\documentclass[fleqn,usenatbib]{mnras}

\usepackage[T1]{fontenc}
\usepackage{ae,aecompl}

\renewcommand{\exp}[1]{\ensuremath{{\rm exp}\left(#1\right)}}

\usepackage{graphicx,subfigure}	
\usepackage{amsmath}	
\usepackage{amssymb}	
\usepackage[figure,figure*]{hypcap}

\title[Robust statistics toward 21~cm EoR signal detection]{Robust statistics toward detection of the 21~cm signal from the Epoch of Reionisation}

\author[Trott et al.]{Cathryn M. Trott$^{1,2, \dagger}$\thanks{cathryn.trott@curtin.edu.au},
Shih Ching Fu$^3$,
S.~G. Murray$^{1,2,6}$,
C.~H. Jordan$^{1,2}$,
\newauthor
J.~L.~B.~Line$^{1,2}$,
N.~Barry$^{13,2}$,
R.~Byrne$^{8}$,
B.~J.~Hazelton$^{8,15}$,
K.~Hasegawa$^{5}$,
\newauthor
R.~Joseph$^{1,2}$,
T.~Kaneuji$^{4}$,
K.~Kubota$^{4}$,
W.~Li$^{16}$,
C.~Lynch$^{1,2}$,
B.~McKinley$^{1,2}$,
\newauthor
D.~A.~Mitchell$^{12}$,
M.~F.~Morales$^{8,2}$,
B.~Pindor$^{13,2}$,
J.~C.~Pober$^{16}$,
M.~Rahimi$^{13}$,
\newauthor
K.~Takahashi$^{4}$,
S.~J.~Tingay$^{1}$,
R.~B.~Wayth$^{1,2}$,
R.~L.~Webster$^{13,2}$,
M.~Wilensky$^{8}$,
\newauthor
J.~S.~B.~Wyithe$^{13,2}$,
S. Yoshiura$^4$,
Q.~Zheng$^{17}$,
M.~Walker$^{1}$\\
$^1$International Centre for Radio Astronomy Research (ICRAR), Curtin University, Bentley WA, Australia\\%
$^2$ARC Centre of Excellence for All Sky Astrophysics in 3 Dimensions (ASTRO 3D), Australia\\%
$^3$EECMS, Curtin University, Bentley, Australia\\%
$^4$Kumamoto University, Japan\\
$^{5}$Graduate School of Science, Nagoya University, Japan\\
$^6$School of Earth and Space Exploration, Arizona State University, Tempe, AZ 85287, USA\\
$^7$Department of Physics, University of Wisconsin--Milwaukee, Milwaukee, WI 53201, USA\\
$^8$Department of Physics, University of Washington, Seattle, WA 98195, USA\\
$^{9}$International Centre for Radio Astronomy Research (ICRAR), University of Western Australia, Crawley, WA 6009, Australia\\
$^{10}$Sydney Institute for Astronomy (SIfA), School of Physics, The University of Sydney, NSW 2006, Australia\\
$^{11}$Western Sydney University, Locked Bag 1797, Penrith South DC, NSW, 1797, Australia\\
$^{12}$CSIRO Astronomy \& Space Science, Australia Telescope National Facility, P.O. Box 76, Epping, NSW 1710, Australia\\
$^{13}$School of Physics, The University of Melbourne, Parkville, VIC 3010, Australia\\
$^{14}$Raman Research Institute, Bangalore 560080, India\\
$^{15}$University of Washington, eScience Institute, Seattle, WA 98195, USA\\
$^{16}$Brown University, Department of Physics, Providence, RI 02912, USA\\
$^{17}$Shanghai Astronomical Observatory, China\\
$^{18}$Curtin Institute of Radio Astronomy, 1 Turner Ave, Technology Park, Bentley, WA 6102, Australia\\
$^\dagger$ARC Future Fellow
}%
\date{Accepted XXX. Received YYY; in original form ZZZ}

\pubyear{2019}

\begin{document}
\label{firstpage}
\pagerange{\pageref{firstpage}--\pageref{lastpage}}
\maketitle

\begin{abstract}
We explore methods for robust estimation of the 21~cm signal from the Epoch of Reionisation (EoR). A Kernel Density Estimator (KDE) is introduced for measuring the spatial temperature fluctuation power spectrum from the EoR. The KDE estimates the underlying probability distribution function of fluctuations as a function of spatial scale, and contains different systematic biases and errors to the typical approach to estimating the fluctuation power spectrum. Extraction of histograms of visibilities allows moments analysis to be used to discriminate foregrounds from 21~cm signal and thermal noise. We use the information available in the histograms, along with the statistical dis-similarity of foregrounds from two independent observing fields, to robustly separate foregrounds from cosmological signal, while making no assumptions about the Gaussianity of the signal. Using two independent observing fields to robustly discriminate signal from foregrounds is crucial for the analysis presented in this paper. We apply the techniques to 13 hours of Murchison Widefield Array (MWA) EoR data over two observing fields. We compare the output to that obtained with a comparative power spectrum estimation method, and demonstrate the reduced foreground contamination using this approach. Using the second moment obtained directly from the KDE distribution functions yields a factor of 2--3 improvement in power for $k < 0.3h$~Mpc$^{-1}$ compared with a matched delay space power estimator, while weighting data by additional statistics does not offer significant improvement beyond that available for thermal noise-only weights.
\end{abstract}

\begin{keywords}
cosmology --- instrumentation: interferometers --- methods: statistical
\end{keywords}



\section{Introduction}
The distribution of line emission signal from high-redshift neutral hydrogen aims to be used as a tracer of physical conditions and structure evolution in the first billion years of the Universe \citep{furlanetto06}. Although weak compared with foreground continuum emitters (brightness temperatures of milliKelvin compared with hundreds of Kelvin), the spatial distribution and redshift evolution of the signal provide insight into the properties of the intergalactic medium, being sensitive to density of gas, temperature of gas, radiation field, and neutral fraction. Its initial detection, and future exploration, are therefore primary experiments of low-frequency radio telescopes, to which the redshifted emission is sensitive, such as the MWA \citep{tingay13_mwasystem,bowman13_mwascience,wayth18}, LOFAR{\footnote[1]{http://www.lofar.org}} \citep{vanhaarlem13}, the Precision Array for Probing the Epoch of Reionization (PAPER){\footnote[2]{http://eor.berkeley.edu}} \citep{parsons10}, and the upcoming Hydrogen Epoch of Reionization Array (HERA){\footnote[3]{http://reionization.org}} \citep{deboer17} and Square Kilometre Array (SKA){\footnote[4]{http://skatelescope.org}} \citep{koopmans15}.

Current instruments have placed upper limits on the spatial power of the brightness temperature distribution of neutral hydrogen gas, but the challenges of complicated instrumentation, precision data calibration, and chromatic foregrounds, have thus far prohibited a detection \citep{thyagarajan15a,liu15,trott16,beardsley16,patil17,paciga11,gehlot18}.

The international community of 21~cm observational scientists is making good progress toward detection of a signal from the Epoch of Reionisation, with recent publications from all major experiments \citep{patil17,gehlot18,cheng18,beardsley16,choudhuri17}. Once a detection is claimed, it is incumbent on the authors to demonstrate that (1) the signal is cosmological, and (2) signal loss has not occurred (i.e., unaccounted loss of cosmological 21~cm power due to the analysis methodology). With many foreground fitting approaches, it is difficult to demonstrate that signal loss has not occurred, and comparison with simulations is limited by the accuracy of the simulation to represent the true (currently unknown) signal structure.

The cosmological signal of brightness temperature fluctuations is expected to be Gaussian-distributed at early times (consistent with the linear phase of structure formation and tracing the matter power spectrum), and evolve to include higher-order terms at later times due to non-linear evolution and the increasing importance of the radiation field \citep{wyithe07}. The bulk of the information is therefore contained in the second moment (variance) of the distribution. Combined with the signal weakness, the power spectrum of temperature fluctuations (normalised variance as a function of spatial scale) therefore presents a natural statistic for initial detection and exploration of the signal as a function of spatial scale, and is a primary data product of the aforementioned experiments. Despite the spherically-averaged (1D) power spectrum providing the largest accumulation of data for the isotropically-distributed signal, often the two-dimensional (2D) power spectrum is used as an intermediate product, whereby the data are retained in angular ($k_\bot$) and line-of-sight ($k_\parallel$) modes. This is to attempt to separate the line emission signal from continuum foregrounds, for which the contamination should be contained in the low $k_\parallel$ modes. The point source population is expected \citep{datta10,trott12,vedantham12,liu15,murray17} and observed \citep{jacobs16,trott16,ali15,beardsley16} to form a wedge-like feature in the low $k_\parallel$ modes, due to the incomplete sampling of a radio interferometer. Despite this `foreground avoidance' technique avoiding >99\% of the foreground emission, chromatic instruments, imprecise calibration, incomplete source models, and limited bandwidth, all combine to allow leakage further into the higher $k_\parallel$ modes, biasing the signal and currently limiting a detection.

The power spectrum is typically estimated via Fourier Transform of the measured interferometric data, coherent accumulation of statistically-equivalent measurements (gridding onto Fourier $uv$-plane), squaring of the averaged accumulated data, and incoherently averaging over annuli of $k_\bot^2 = u^2+v^2$ \citep[e.g., ][]{datta10}. This provides an estimator of the variance for a given spatial scale. In the presence of Gaussian-distributed noise and signal, this provides an unbiased estimate of the power. In the presence of strong foregrounds, which contribute non-Gaussian signal that is not statistically-representative of the data (e.g., a bright extended source at a single sky location), this method captures all of that excess signal, yielding bias in the output power spectrum. This motivates the use of Kernel Density Estimation method to provide a different path to obtaining the spatial power, which may be more robust to foreground-induced outlier data and allow for more ready access to the Gaussian-distributed signal.

KDEs were developed more than half a century ago, with credit for their motivation and construction to \citet{rosenblatt56} and \citet{parzen62}. When trying to estimate the population distribution from data, one attempts to reconstruct the population properties via estimation of the sample distribution. A direct histogram of measured values has the potential to lead to a discretized output, with noisy measurements yielding bias in the estimates. For any estimation procedure, larger amounts of identically-distributed data will always yield a more precise and accurate representation of the underlying distribution.

The KDE attempts to address the discretization problem by including each measurement in the estimate as a convolution of the data value with a known, pre-defined kernel function. This smoothing has the benefit of naturally reducing discretization, but has the potential to overestimate the distribution width (through a kernel that is too broad) or bias the result (through a kernel function that is not well-matched to the underlying distribution function). For our correlation data (the real and imaginary components of the measured visibilities), each 8-second sample is highly thermal-noise dominated (even in the presence of residual foregrounds, after bright foreground removal), providing a natural choice of kernel as the Gaussian distribution.

\section{Methods}\label{sec:methods}
\subsection{Constructing a Kernel Density Estimator}
A KDE aims to construct an estimate for the distribution function of identically distributed data, using the data themselves. The kernel is the underlying functional form for determining the contribution of a given measurement to the overall estimate. For an estimate with minimal bias, the kernel is best to be matched to the true underlying distribution, although this is often unknown (hence the use of KDE).

For complex-valued visibility data from a radio interferometer with high spectral and temporal resolution (as is the case for EoR data), each data sample will be heavily thermal noise-dominated, rendering a Gaussian kernel to be a natural choice for our purposes. Moreover, the use of a KDE to construct the power spectrum necessitates a Gaussian kernel from which the variance can be equated to the power.

For a set of $N$ identically- and independently-distributed (iid) data, $x_j$, with mean value $\bar{x}$, an estimate of their distribution can be found via:
\begin{equation}
\hat{f}(x) = \frac{1}{Nh}\displaystyle\sum_{j=1}^N K\left( \frac{x_j-\bar{x}}{h} \right),
\end{equation}
where the summation extends over the iid data, $K()$ denotes the compact kernel function, and $h$ is a scaling of the breadth of the kernel, for which an optimal value for Gaussian-distributed data is found to be \citep{silverman86}:
\begin{equation}
h \approx 1.06\sigma(N)^{-1/5},
\end{equation}
where $\sigma$ is the standard deviation of the data. For radio interferometric visibilities, for which the real and imaginary components contain an equal share of 21-cm signal and Gaussian noise measured in Janskys, $S$:
\begin{equation}
\hat{f}(S_i) = \frac{1}{N_{i}h}\displaystyle\sum_{j=1}^{N_i} K\left( \frac{S_{ij}(k_\bot,k_\parallel)-S}{h} \right),
\end{equation}
with
\begin{equation}
K\left( \frac{S_{i}(k_\bot,k_\parallel)-S}{h} \right) = \exp{-\frac{(S_{i}(k_\bot,k_\parallel)-S)^2}{2h^2}},
\end{equation}
and for $i \in [k_\bot,k_\parallel]$. We then connect the variance to the power and equate,
\begin{equation}
A_i\exp{-\frac{(S_{i}(k_\bot,k_\parallel)-S)^2}{2P_S(k_\bot,k_\parallel)}} = \displaystyle\sum_{j=1}^{N_i} \exp{-\frac{(S_{ij}(k_\bot,k_\parallel)-S)^2}{2h^2}},
\end{equation}
where $P_S(k_\bot,k_\parallel)$ is the power measured in units of Jy$^2$, and $A$ is an amplitude that depends on the number of visibilities contributing to cell $i$, and is unused except to estimate the noise uncertainty. The optimal value for $h$ is found by estimating the number of visibilities contributing to each spatial mode, and is computed from the known baseline distribution and a pre-defined amount of input data (total observing time). The weak dependence of $h$ on $N_i$ allows for an approximation of this number to be sufficient. For an ultimate EoR detection, the kernel gridding size needs to be smaller than the expected 21~cm signal strength, otherwise the discretization of the grid will yield an unphysical upper limit. This choice is discussed in Section \ref{sec:data}.

\subsubsection{Limitations}
Unlike a direct variance estimator, which squares averaged equivalent data (i.e., data that sample the same sky signal), the KDE relies on the data being \textit{statistically}-equivalent, and providing a representative sample of the full distribution (i.e., for $uv$ data, while each datum provides a different statistical realisation of the thermal noise, each $u,v$ point in an annulus of constant $k_\bot$ provides a different statistical realisation of the 21-cm signal and the foregrounds). Therefore, a single sampled $u,v$ location for a given $k_\bot$ will not provide the underlying distribution function for the 21-cm signal, and will yield a biased estimator of the variance. Therefore, the success of the KDE relies on complete and balanced $uv$-coverage for a given $k_\bot^2 = u^2+v^2$ annulus and benefits from accumulation of data from multiple observing fields. By construction, the KDE must use information from different baseline vectors (but of similar length). This is because these different vectors capture different 21cm statistical realisations, allowing the KDE-histogram to be representative of the statistics of the signal. The sample variance can be used to estimate which modes contain enough independent samples to be statistically reliable. 

The second limitation comes from the thermal noise uncertainty reduction with quantity of data. While the gridded visibility data allows for coherent averaging of statistically-identical data before incoherent averaging to the spherical power spectrum, the KDE accumulates information incoherently. Thus, instead of gaining by $N$ samples for these equivalent data, the KDE always gains by $\sqrt{N}$. This does play a role in requiring more data to be accumulated to reach a thermal noise level, however much of the averaging for a spherical 1D power spectrum is incoherent, and thermal noise reduction is not the limiting factor for current experiments. If the KDE can improve the systematic errors sufficiently, then this will still provide benefit over stochastic error reduction.

\subsubsection{Extracting the Cross Power Spectrum from a KDE}
Building the data distribution directly from the \textit{iid} real and imaginary components of the complex-valued visibilities allows use of a Gaussian kernel for the estimator. The kernel size parameter, $h$, is determined by the expected thermal noise for a given visibility and the number of observations accumulated for the final estimate. The variance estimated from such a constructed distribution will yield the thermal noise \textit{power}, just as the power spectrum constructed from squaring of visibilities (the auto power spectrum) will be dominated by the thermal noise power. To remove the approximated (but unknown) noise power, the \textit{cross} power spectrum can be constructed \citep[such as is used by][]{trott16,beardsley16,barry19}, by cross-multiplying matched data (e.g., closely-spaced interleaved time samples or frequency samples). The cross power spectrum contains no noise bias, but only noise uncertainty, due to the sample variance remaining from the differing realisations of the noise. In a similar fashion, the cross KDE power spectrum can be constructed. The cross power, formed from two identically-distributed sets of complex-valued data $\vec{x}_1$ and $\vec{x}_2$, is given by:
\begin{eqnarray}
P &=& \frac{1}{2}\left(\vec{x}_1^\ast \vec{x}_2 + \vec{x}_2^\ast \vec{x}_1 \right)\nonumber\\
&=& \frac{1}{2}\left((\vec{x}_1+\vec{x}_2)^\ast(\vec{x}_1+\vec{x}_2) - (\vec{x}_1-\vec{x}_2)^\ast(\vec{x}_1-\vec{x}_2) \right)\nonumber\\
\label{eqn:diff}
&=& P_{\rm tot} - P_{\rm diff}
\end{eqnarray}
and is therefore equivalent to the difference in the summed auto power and the differenced auto power. The cross KDE power spectrum can be constructed similarly by summing (`Totals') and differencing (`Differences') time-interleaved datasets, estimating their individual variances from the KD estimates, and taking the difference. In addition to providing the cross power spectrum, the difference power, $P_{\rm diff}$, contains an estimate of the noise power. Throughout, we refer to these two sets of data as the Totals and Diffs.

Each visibility is weighted by the data content (the visibility weights, which are a product of the channel bandwidth and temporal resolution), and a spectral taper is applied prior to spectral Fourier Transform ($\nu\rightarrow\eta$) to reduce spectral leakage. For both we employ a Blackman-Harris taper.

In this work, we employ a delay spectrum approach to computing the power spectral density for a given scale \citep{parsons10}. The delay spectrum Fourier Transforms directly over the spectral channels for a given baselines's visibility set, and therefore is not a strictly line-of-sight transform (due to the $|u|$ value for a baseline changing with frequency). The delay transform is closely related to the pure line-of-sight transform, but is not parallel, except for the $k_\bot=0$ mode. For non-zero $k_\bot$, it transforms along a direction with some angle to the line-of-sight, where the angle increases with $k$. I.e., it mixes angular modes with line-of-sight modes, thereby producing a non-linear mapping to $k_\bot-k_\parallel$ space. In addition, a direct frequency transform of visibilities neglects any effects of the convolution of the true signal with the instrument primary beam, and therefore does not easily capture correlations between bins. As such, the delay spectrum cannot directly be interpreted cosmologically. However, for this work, and for short baselines of relevance here, it provides a computationally simple approach to forming KDE estimates of the power. Throughout, each output in this work uses the same underlying delay spectrum approach, and therefore the results are internally consistent.

Specifically, to form the KDE histograms, we take the following steps:
\begin{enumerate}
    \item For each snapshot observation, the 384 spectral channels for each of the 8128 MWA baselines is combined into two datasets: sums of time-contiguous visiiblities, and differences;
    \item Each set is separately Fourier Transformed to compute its contribution to a given $\eta$ mode (i.e., $V(\eta) = \displaystyle\sum_j V_j(\nu) \exp{-2\pi{i}\nu_j\eta}$ for baseline $j$);
    \item The 8128 baselines of data from that snapshot are then gridded onto the KDE with a Gaussian kernel, where the individual baselines are each assigned to the KDE for a single angular mode ($k_\bot$ with $\Delta{u} = 2.5\lambda$) according to their $u,v,w$ values at the lowest frequency channel. No beam gridding is performed;
    \item The next snapshot is read and the same procedure is applied. Those data are gridded onto the same KDE distributions.
\end{enumerate}
At the end of this process, there is a KDE histogram distribution for each $\eta$ and $k_\bot$ cell. The datasets used here contain $\simeq$1 billion visibilities in each histogram. The comparison delay space power spectrum uses the same approach, but the summed and differenced data for each baseline and snapshot are squared to form power, before combining incoherently with the remaining baselines and snapshots (i.e., each summed or differenced visibility, after Fourier Transforming, is squared, and the combined with other visibilities of the same angular scale using a weighted average). This provides a direct comparison between the two approaches.  As before, the crosspower is extracted by subtracting the power in the differences from the power in the totals, as in Equation \ref{eqn:diff}.

After forming the KD estimate for each ($k_\bot,k_\parallel$) cell, the power needs to be extracted by fitting a Gaussian distribution and using the second moment, or, by directly computing the sample variance from the histogrammed data.

\section{Interrogating the density estimates}
Analysis of the KDE-derived histogram distribution functions can be approached in a number of ways, guided by their form. In this work, we use several approaches to assess their effectiveness:
\begin{enumerate}
    \item Moments analysis: compute the first four moments of the real and imaginary component histograms to assess Gaussianity as a function of angular and LOS scale;
    \item Phase analysis of moments: compute the phase of the first four moments for each observing field;
    \item Earth Mover's Distance (statistical similarity): compute the dis-similarity of the histograms for each observing field as a function of angular and LOS scale.
\end{enumerate}
Approach (1) has the potential to destroy non-Gaussian EoR signal, while (2) and (3) can preserve it. Each output of these approaches is used to weight data in an attempt to improve the power spectrum, and the results are compared in Section \ref{sec:data}.

\subsection{Moments Analysis}
If the signals were expected to decompose into summed Gaussians, then fitting multiple Gaussians and extracting parameters would be useful. However, this approach cannot guarantee that 21cm signal is not mixed into foreground signal. Instead, we can be motivated by the expected shape of the signal: \citet{wyithe07} study the theoretical brightness temperature distribution functions for the cosmological signal as a function of scale and redshift. They find that the signal is primarily Gaussian, but exhibits non-zero skewness (third moment) for some scales and redshifts. This skewness does not exceed 0.01. There is no fourth-moment (kurtosis) present. \citep[More recent analysis by][shows similar results.]{majumdar18} To study the observed visibility distributions, we can perform moments analysis directly on the KDE outputs\footnote{The first four moments are:
\begin{itemize}
    \item Mean = $\bar{x} = \frac{1}{N} \displaystyle\sum_{i=1}^N x_i$
    \item Variance = $\frac{1}{N-1} \displaystyle\sum_{i=1}^N (x_i-\bar{x})^2$
    \item Skewness = $\frac{1}{N} \displaystyle\sum_{i=1}^N \left(\frac{(x_i-\bar{x})}{\sqrt{{\rm Var}}}\right)^3$
    \item Kurtosis = $\frac{1}{N} \displaystyle\sum_{i=1}^N \left(\frac{(x_i-\bar{x})}{\sqrt{{\rm Var}}}\right)^4 - 3$
\end{itemize}}.
Large mean and skewness values, and non-zero kurtosis, suggest foreground contamination and can guide the next steps in the analysis.

\subsection{Phase analysis of moments}
The moments are computed for histograms of the real and imaginary components of the visibilities. This is distinct from typical power spectrum analysis, where the visibilities are gridded coherently (with phase) and then squared to form the power (i.e., phase information is lost). Although we expect that the statistical properties of the 21~cm signal from the EoR are the same between fields (up to cosmic variance), the phase of the signal for any given vector coherent $\vec{k}_\bot$ mode (a single cell in the $uv$-plane) can differ. However, when computing the histogram of visibilities for all angular modes satisfying $|\vec{k}_\bot| = k_\bot$, each cell contributes a random phase and the real and imaginary component histograms are expected to be similarly distributed. Departure from similarity of the real and imaginary histograms can occur for: (1) sufficiently few cells to provide a statistical sample; or (2) strong, non-stochastic signals imprinting direct structure. The former case is handled by omitting modes for which there are few independent samples (small $k_\bot$). It is the latter case that incorporates the effect of strong foregrounds that leave structure in the sky. E.g., strong extended sources (Fornax A), and the Galactic Plane.

The phase of a given visibility cannot be extracted from the data, but the phase of the KDE-derived moments for each $k_\bot-k_\parallel$ cell can be used to study parts of the parameter space that display phase significantly different from $\pi/4$ (equal share for the real and imaginary components). We therefore can study the phase for each moment using the real and imaginary component values for the moments in each $k_\bot-k_\parallel$ cell such that:
\begin{equation}
    \phi_j(k_\bot,k_\parallel) = \text{atan}\left(\frac{Im(M_j)}{Re(M_j)}\right) - \frac{\pi}{4}.
\end{equation}
for moment $M_j$, where this is performed from the individual moments values for each $k_\bot$ annulus histogram, at each $k_\parallel$ line-of-sight cell.

\subsection{Earth Mover's Distance}
The different foregrounds present in two distinct observing fields will imprint different structures into the KDE distribution functions. We want to use this dis-similarity to discriminate statistically dis-similar foregrounds from statistically similar cosmological signal.
Therefore, the lack of similarity between the distribution functions from two observing fields for the same $k$-mode is an indicator of foregrounds. Statistically dis-similar histograms suggest foreground contamination because the EoR signal is expected to be isotropic and homogeneous, and therefore is statistically similar across the sky. We use the Earth Mover's Distance \citep[EMD; also known as the Wasserstein Metric]{olkin82} as a measure of the similarity of the two distributions measured from the two fields. Large values of the EMD (exceeding that expected from thermal noise uncertainty) can be used to identify a component of foreground contamination; statistically-similar foregrounds are preserved in this analysis. \textit{Crucially, statistically-similar, but non-Gaussian, 21~cm signal is also preserved, allowing EMD analysis to provide a robust measure of some fraction of the foregrounds without risk of signal loss}. For two histograms, one from each observing field, with cumulative distribution functions (CDFs) given by $\Phi_1(x)$ and $\Phi_2(x)$, the EMD is given by:
\begin{equation}
    \rm{EMD} = \displaystyle\sum_i |\Phi_1(x_i) - \Phi_2(x_i)|,
\end{equation}
where $i$ indexes over the flux density values of the KDE-derived CDFs. In all cases, the EMD is computed for a specific $k_\bot-k_\parallel$ cell.
The EMD is bounded from below by zero, and above by a value dependent on the difference between the mean values for each distribution.

Error analysis can be used to determine the expected EMD value based on thermal noise uncertainty for each distribution (computed individually for the dataset from each observing field), a non-negative value for finite datasets. 
Propagating the error for the CDF and then summing to account for the absolute-value sum, we find that:
\begin{equation}
    \langle{\rm EMD}\rangle = \sqrt{\frac{\Delta\sigma^2}{2\pi\sigma^6}} \sqrt{\displaystyle\sum_i \left( \int_{-\infty}^{x_i} t^2 \exp{-t^2/\sigma^2}\right)^2} ~dt,
\end{equation}
where $\sigma$ and $\Delta\sigma$ are the expected noise power and noise uncertainty respectively for each dataset (the latter limited by the kernel resolution). With the dataset used, we expect:
\begin{equation}
    \langle{\rm EMD}\rangle \simeq 1.5,
    \label{eqn:emd}
\end{equation}
where we emphasise that this is a non-negative quantity.

\section{Data}
\label{sec:data}
\begin{figure}
\includegraphics[width=0.45\textwidth]{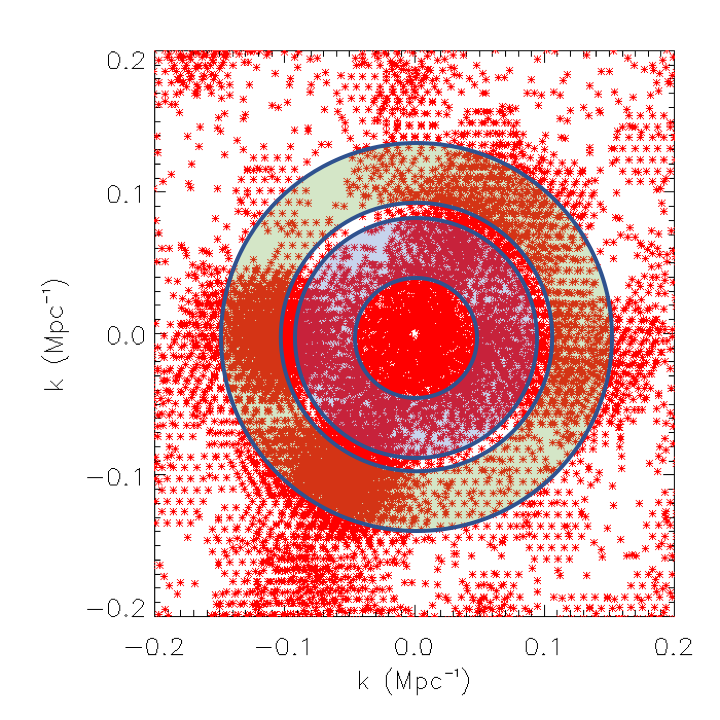}
\caption{$uv$-coverage of a zenith snapshot at a single frequency of 187~MHz with the core and two symmetric hexagons of the MWA Phase II array. The coverage demonstrates the angular symmetry required for the KDE to yield a statistically-representative distribution of the 21-cm signal. The two annuli show diagrammatically examples of regions from which the statistical histograms for the KDE are built. Regions extend from the centre of the $uv$ plane to radii at which the coverage does not adequately sample the different angles.}
\label{fig:uv}
\end{figure}

Data from the 2016 and 2017 EoR seasons of the MWA are used, across two distinct observing fields. The use of both fields to discriminate signal from foregrounds is crucial for the analysis presented in this paper. These data are observed with Phase II of the array, in which the 128 individual tiles form a compact configuration, and include two hexagonal subarrays optimised for EoR science \citep{wayth18}. These data are chosen primarily because the MWA core, and two hexagonal subarrays, provide the angular symmetry in $uv$-coverage required for the KDE to adequately sample different statistical realisations of the 21-cm signal. Figure \ref{fig:uv} shows the inner $uv$-coverage for a single frequency zenith snapshot at 187~MHz. Each point is a single baseline, and the influence of the symmetry of the hexagons is apparent.
Sample variance can further be improved by including data from multiple observing fields in each KD estimation, and sample variance estimates are used in this work to ensure that these modes have well-sampled statistics in order for the KDE to yield reliable distributions.

For this experiment, we employ EoR high-band data, comprising 384 80~kHz spectral channels over 30.72~MHz of contiguous bandwidth in range 167--197~MHz ($z=7.5-6.2$). These data are observed in two of the primary MWA observing fields, EoR0 (RA=0h, Dec=--27deg.) and EoR1 (RA=4h, Dec=--27deg.) \citep[details described in][]{jacobs16}. We process 400 2-minute observations, with 8-second temporal interleaving of two matched datasets, observed in 2016 August -- 2017 December. The data are chosen such that they show low ionospheric activity \citep{jordan17} and no obvious calibration artifacts in the time-interleaved differenced data. The data are calibrated (direction-independent and direction-dependent) with the MWA Real-Time System \citep[RTS,][]{mitchell08} and 1000 sources are peeled from the calibrated visibilities. This is the standard processing applied to MWA EoR data. These calibrated and peeled visibilities are used as input to both the KDE and direct delay transform pipelines (note that this is different to the standard CHIPS pipeline used for MWA EoR analysis, \citet{trott16}).

The KDE approach is optimised for a flux density resolution and kernel characteristic size, $h$, commensurate with the noise level in the data. For each of the real and imaginary components of the visibilities, the expected noise level for a single 8~second visibility and 30.72~MHz bandwidth (after Fourier Transforming along the frequency dimension) is:
\begin{equation}
    \Delta{V} = \frac{{\rm SEFD}}{\sqrt{2\Delta{t}\Delta{\nu}}} \simeq 1.3~{\rm Jy}.
\end{equation}
With one billion visibilities in the dataset for an average large angular scale, this yields an optimal scaling of:
\begin{equation}
    h \simeq 1.06\sigma (N)^{-1/5} = 0.02~{\rm Jy}.
\end{equation}
In this work, we use $\delta{V}=0.01$~Jy to resolve (bin) the KDE, but note that in general this resolution would not be sufficient to discriminate the EoR signal (but is adequate for the data volume used herein). A higher resolution can be used when more data are included, at the expense of computational load.

\section{Results}
\begin{figure}
\includegraphics[width=0.5\textwidth]{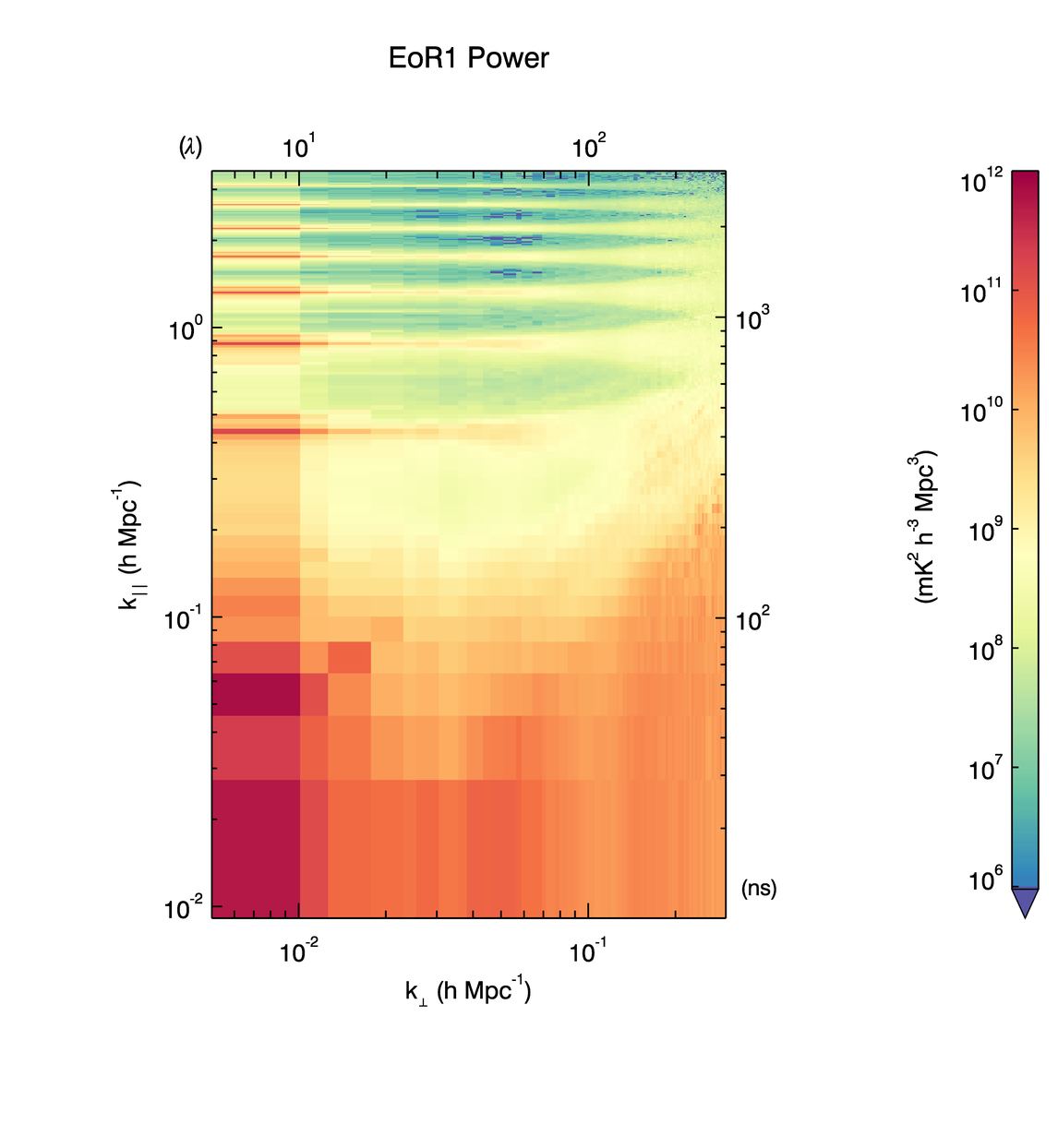}
\caption{Power spectrum of data from the EoR1 field, processed through a delay space estimator.}
\label{fig:eor1_ps}
\end{figure}

For a point of comparison, the datasets are processed through a delay-transform procedure, as outlined in Section \ref{sec:methods}. The delay power spectrum for EoR1 is displayed in Figure \ref{fig:eor1_ps}.
From this delay power spectrum, there are a few key features that will be of relevance for the analysis to follow:
\begin{itemize}
    \item The wedge-like signature of foregrounds extending from low $k_\bot-k_\parallel$ to $k_\bot \simeq 0.3h$~Mpc$^{-1}$, $k_\parallel \simeq 0.3h$~Mpc$^{-1}$;
    \item Bands of foreground power parallel to the diagonal wedge, due to the primary beam shape;
    \item $k_\parallel$ harmonics of the $k_\parallel=0$ mode due to regular missing spectral channels;
    \item Excess power beyond the wedge in the `EoR Window' ($k_\parallel < 0.45h$~Mpc$^{-1}$), due to leakage of foreground power into larger LOS modes.
\end{itemize}

\subsection{KDE Power Spectrum}
\begin{figure}
\includegraphics[width=0.5\textwidth]{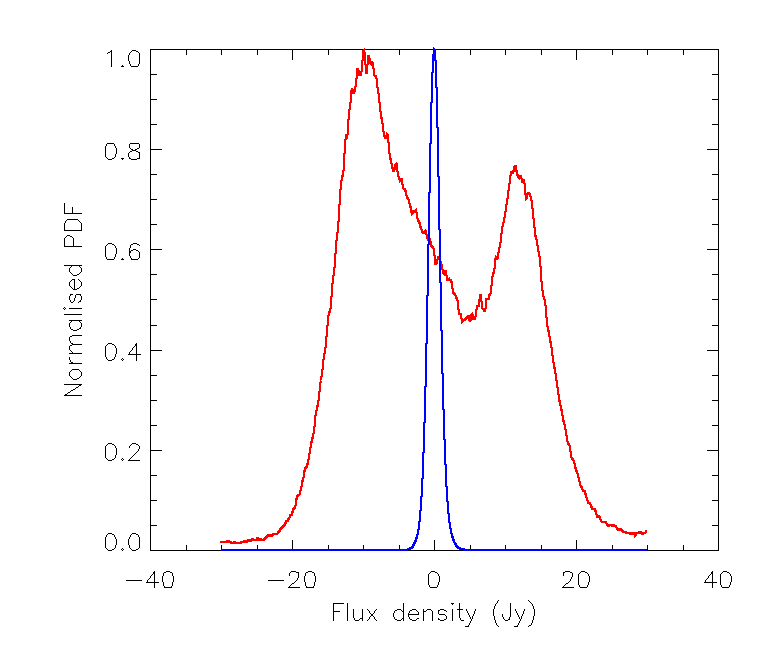}
\caption{Example histograms of the totals (red) and difference (blue) visibilities for a small $k_\bot-k_\parallel$ mode, where foregrounds are expected to dominate. Here, the kernel scale, $h=0.02$~Jy, and is therefore too small to show on the plot.}
\label{fig:histo_small}
\end{figure}

\begin{figure}
\includegraphics[width=0.5\textwidth]{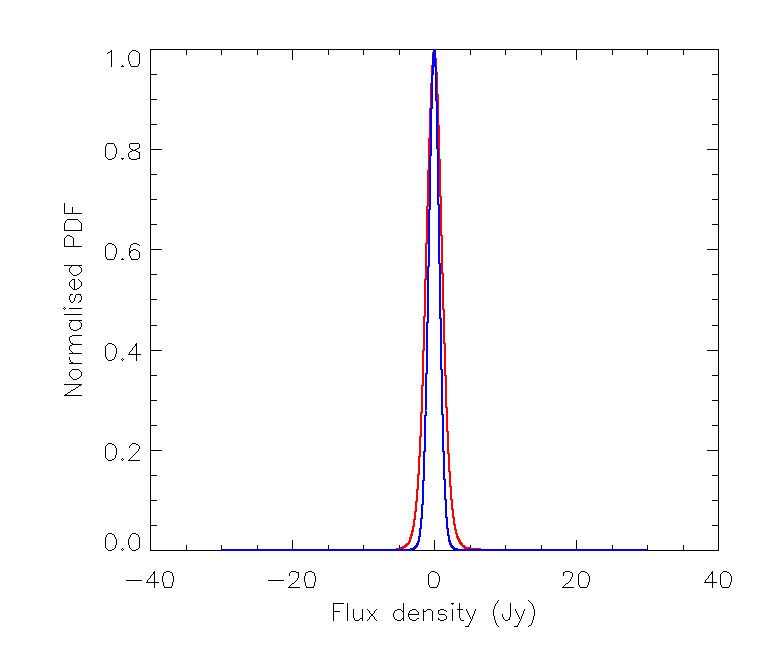}
\caption{Example histograms of the Totals (red) and Differences (blue) visibilities for a large $k_\bot-k_\parallel$ mode within the EoR Window, where foregrounds are not expected to dominate, but do show the small power increase that is leaked power visible in this region in Figure \ref{fig:eor1_ps}.}
\label{fig:histo_large}
\end{figure}

The histograms constructed from the calibrated data vary markedly across $k$-space. At large scales ($k_\bot,k_\parallel < 0.1h$~Mpc$^{-1}$), where foregrounds are known to dominate, the totals visibilities exhibit broad, non-Gaussian histograms. Difference visibilities from the same modes are narrower, indicative of being mostly noise-like, but do exhibit some non-Gaussianity from leaked foreground power (Figure \ref{fig:histo_small}).
At scall scales ($k_\bot \simeq 0.05h$~Mpc$^{-1} ,k_\parallel \simeq 0.1h$~Mpc$^{-1}$), the totals and difference histograms are more similar (Figure \ref{fig:histo_large}). However, even here the Totals distribution is broader than the Differences and this increased power translates to the leaked foreground power entering the EoR Window in Figure \ref{fig:eor1_ps}. Despite this region generally having 3--4 orders of magnitude less foreground power than the wedge, it still exceeds the noise power, and 21~cm power, in this region. It is this additional leakage that we want to identify and remove.
The difference in breadth of the two distributions yields the 21~cm power (in the absence of foregrounds). The width of the Difference visibilities reflects the noise power for a single Fourier Transformed visibility.

\begin{figure*}
\includegraphics[width=0.45\textwidth]{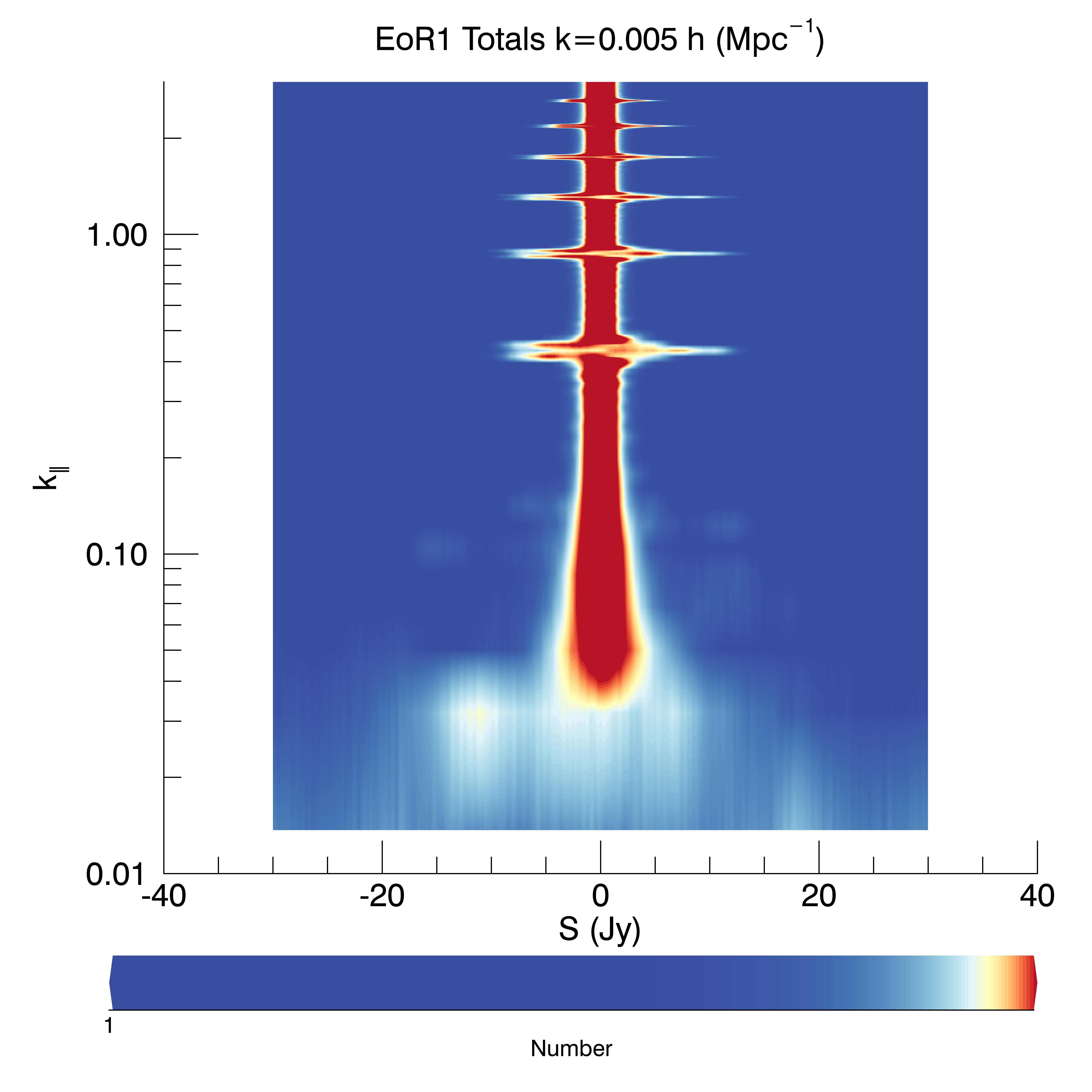}{(a)}
\includegraphics[width=0.45\textwidth]{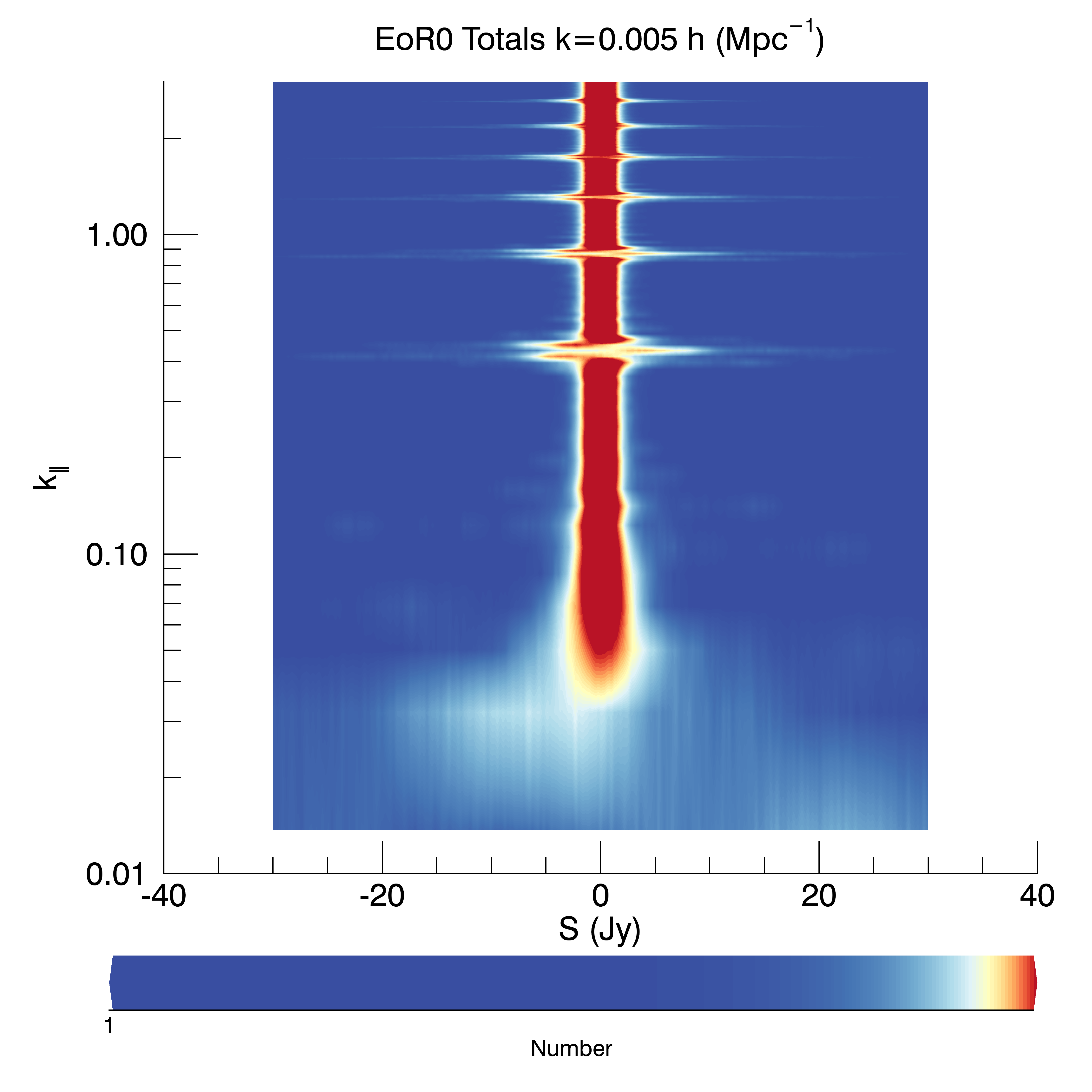}{(b)}\\
\includegraphics[width=0.45\textwidth]{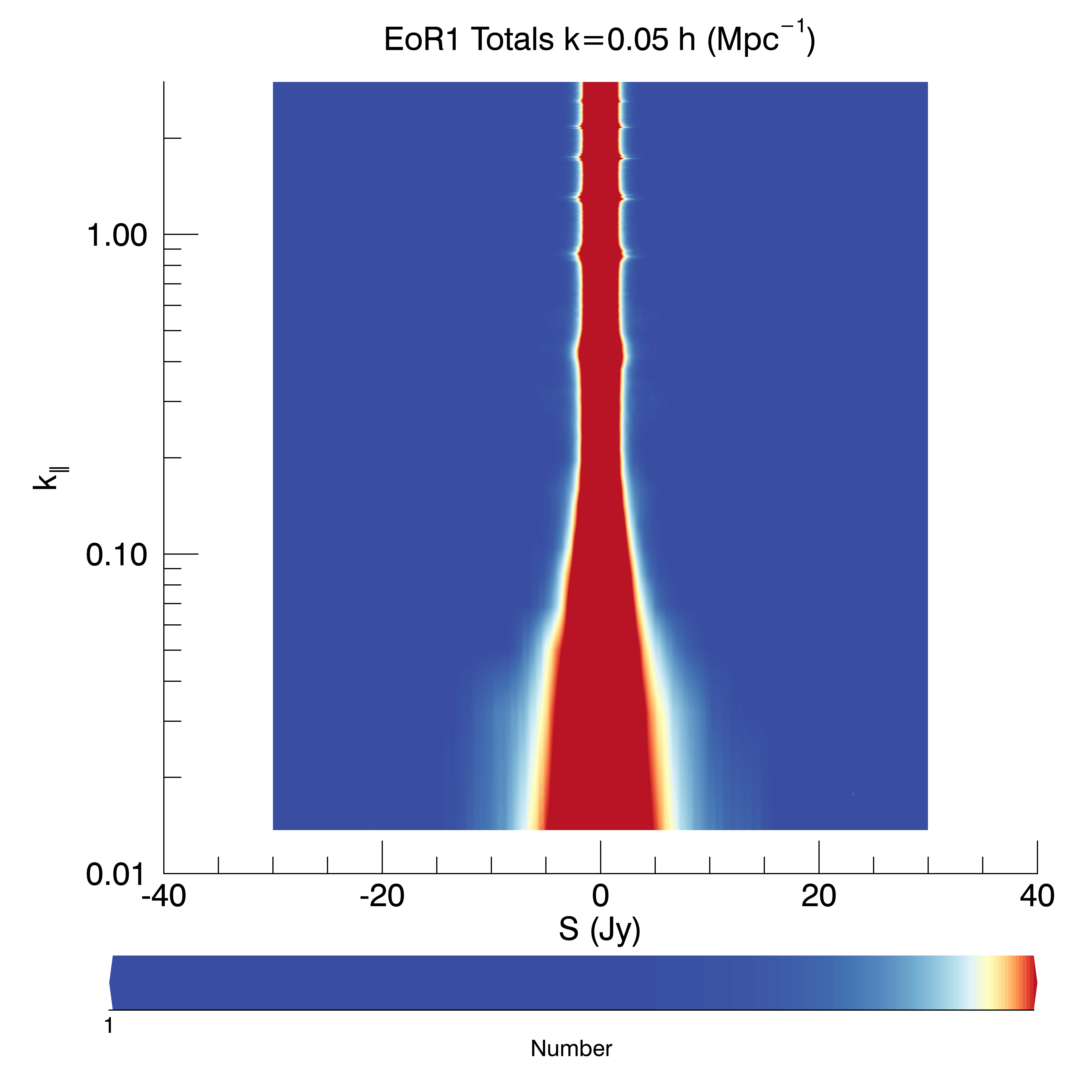}{(c)}
\includegraphics[width=0.45\textwidth]{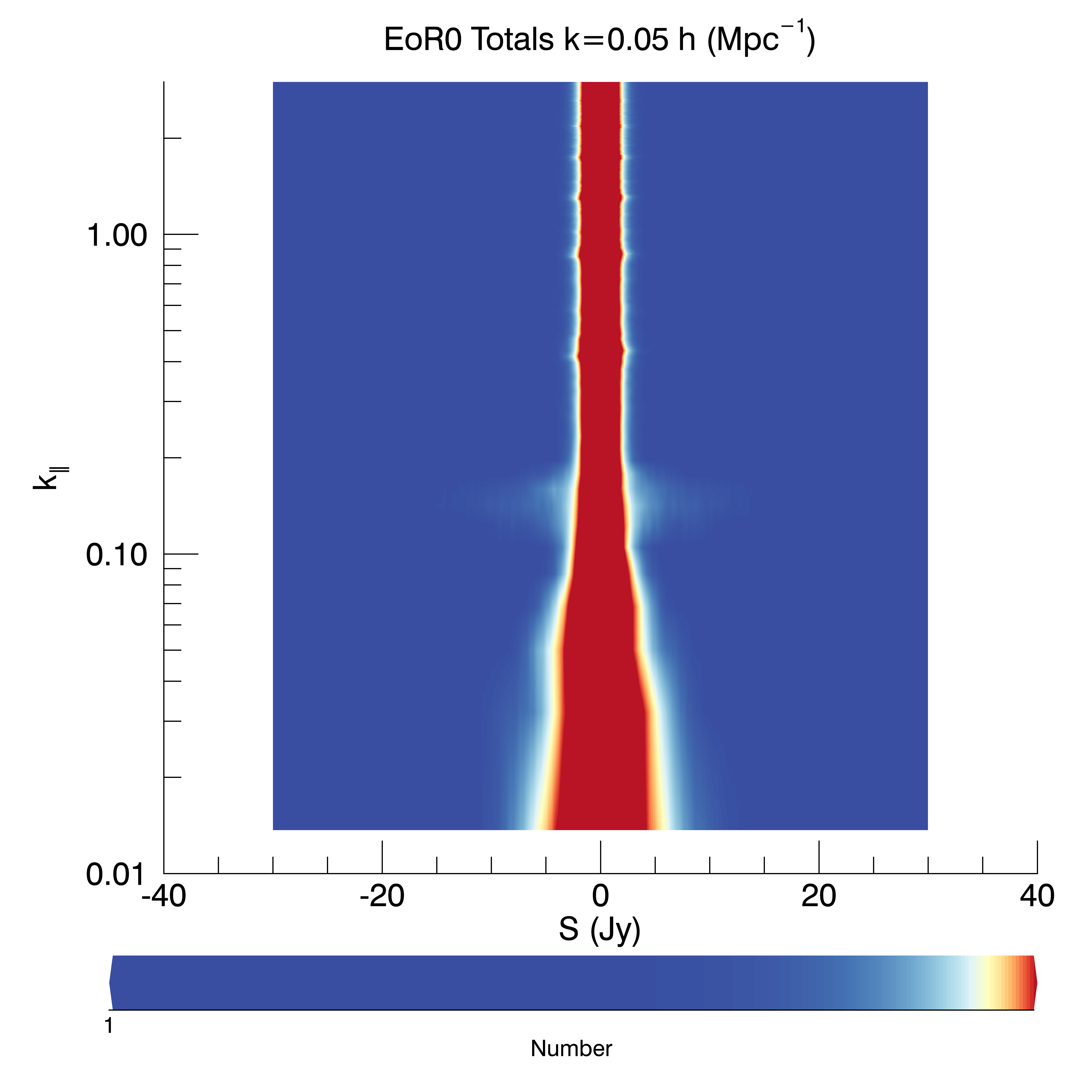}{(d)}
\caption{Heat maps {of the real parts} of the visibilities (abscissa: flux density; ordinate: $k_\parallel$) of density estimates from 13 hours of data. (a, b) Total visibilities (sum over adjacent interleaved time samples per baseline) at $k_\bot=0.005h$~Mpc$^{-1}$ for both fields. (c, d) Same, but for smaller scales, $k_\bot=0.05h$~Mpc$^{-1}$. All colour bars have the same scale, {corresponding to 0.1\% (white), 1\% (yellow), 10\% (orange), 50\% (red) of maximum.}}
\label{fig:teststat2}
\end{figure*}

\begin{figure*}
\includegraphics[width=0.45\textwidth]{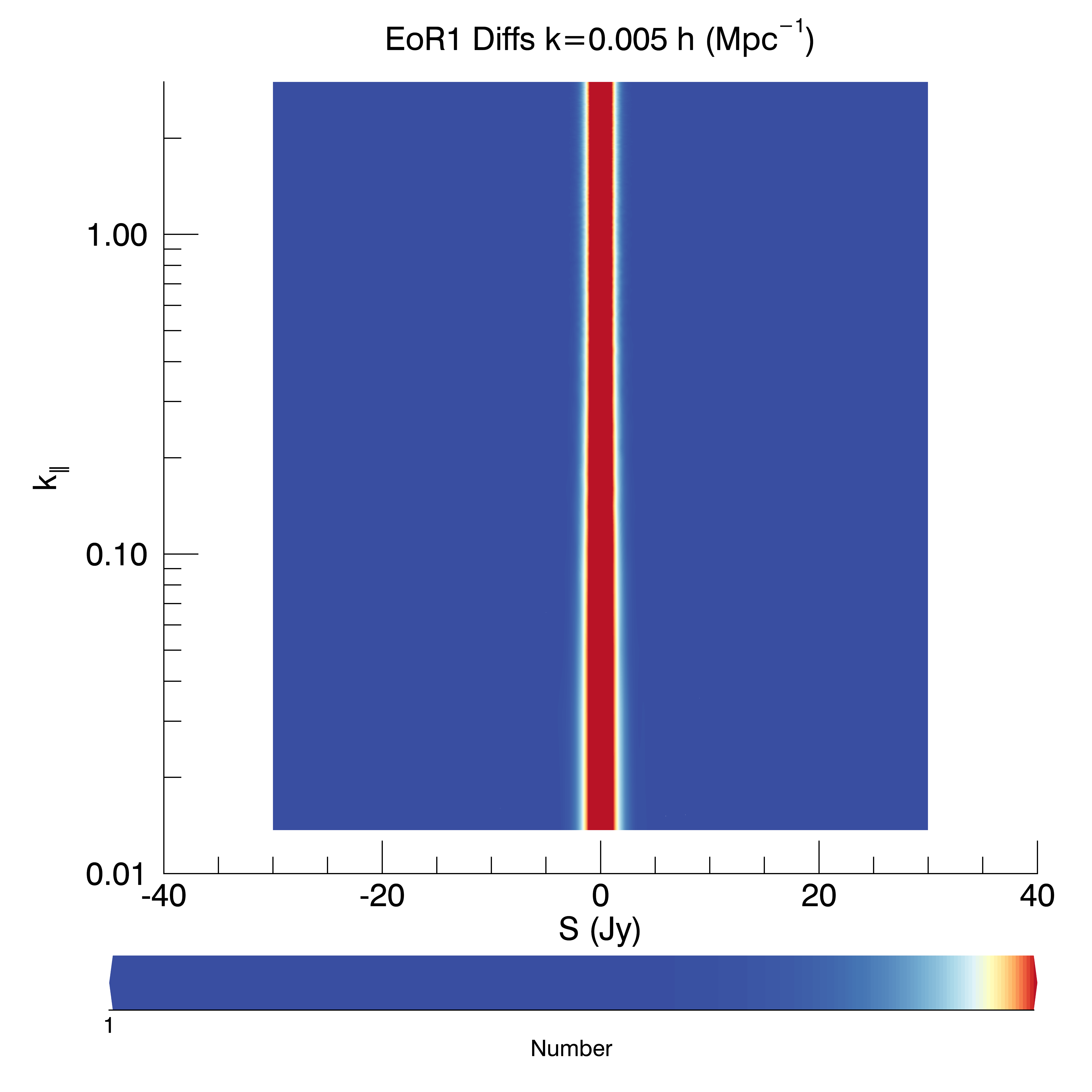}{(a)}
\includegraphics[width=0.45\textwidth]{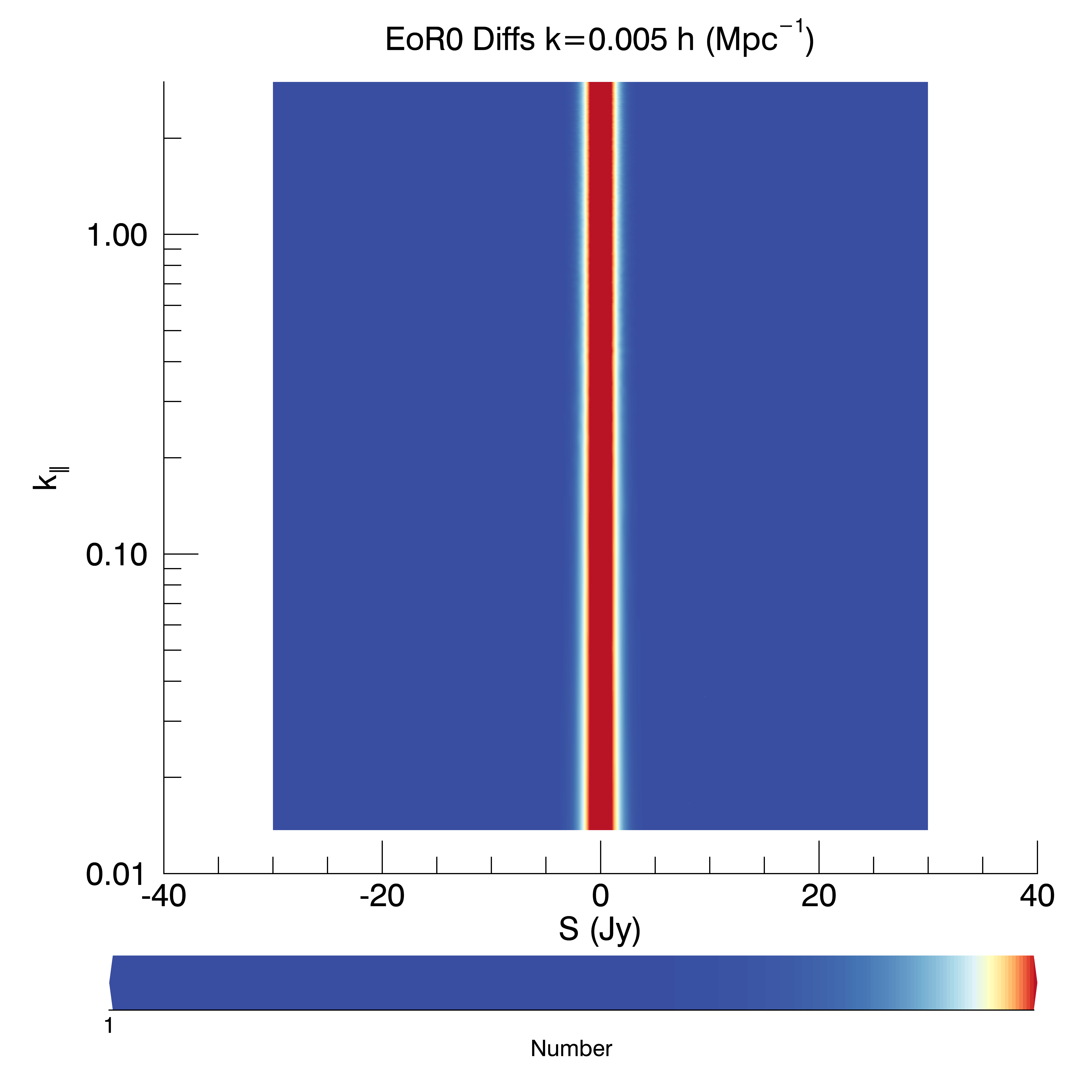}{(b)}\\
\includegraphics[width=0.45\textwidth]{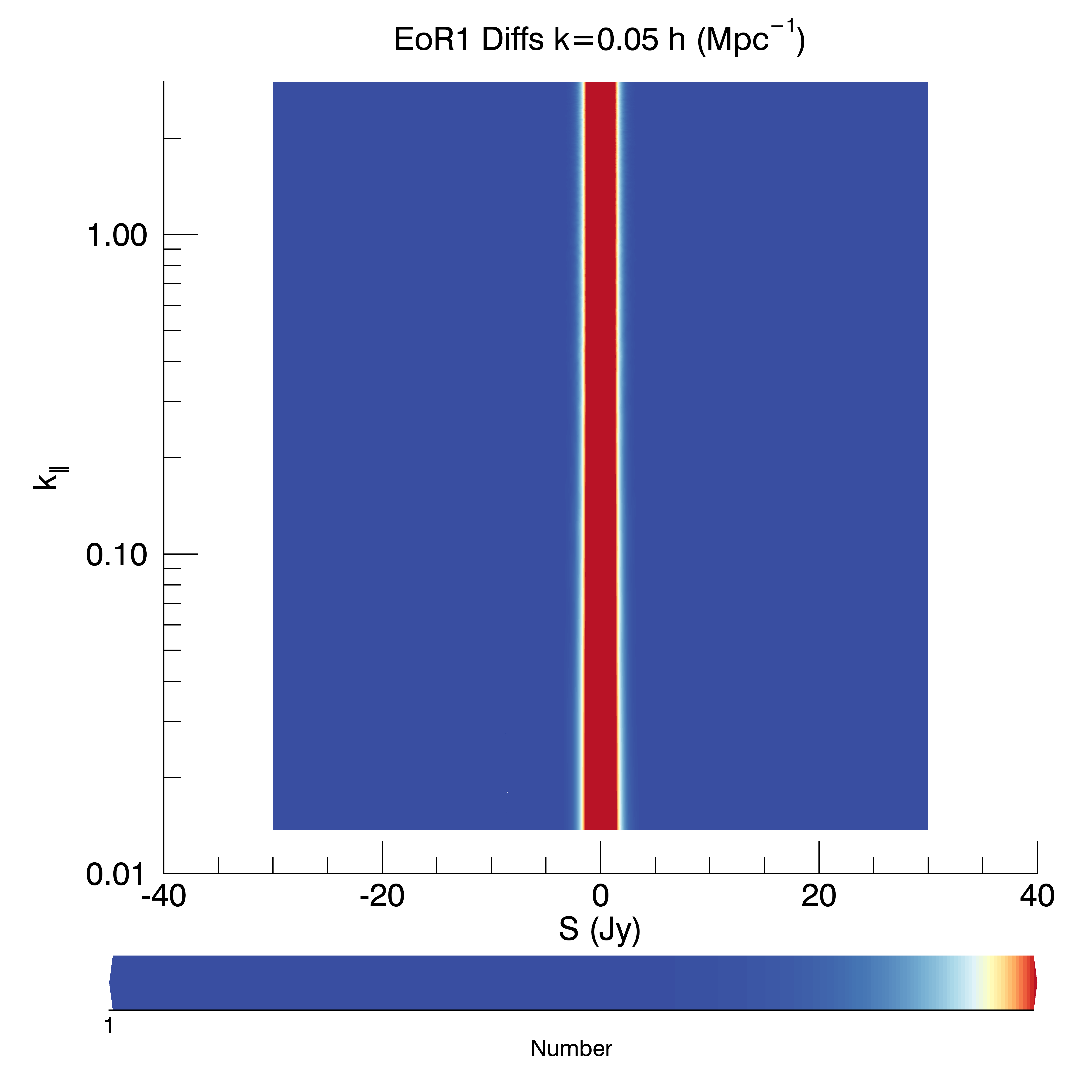}{(c)}
\includegraphics[width=0.45\textwidth]{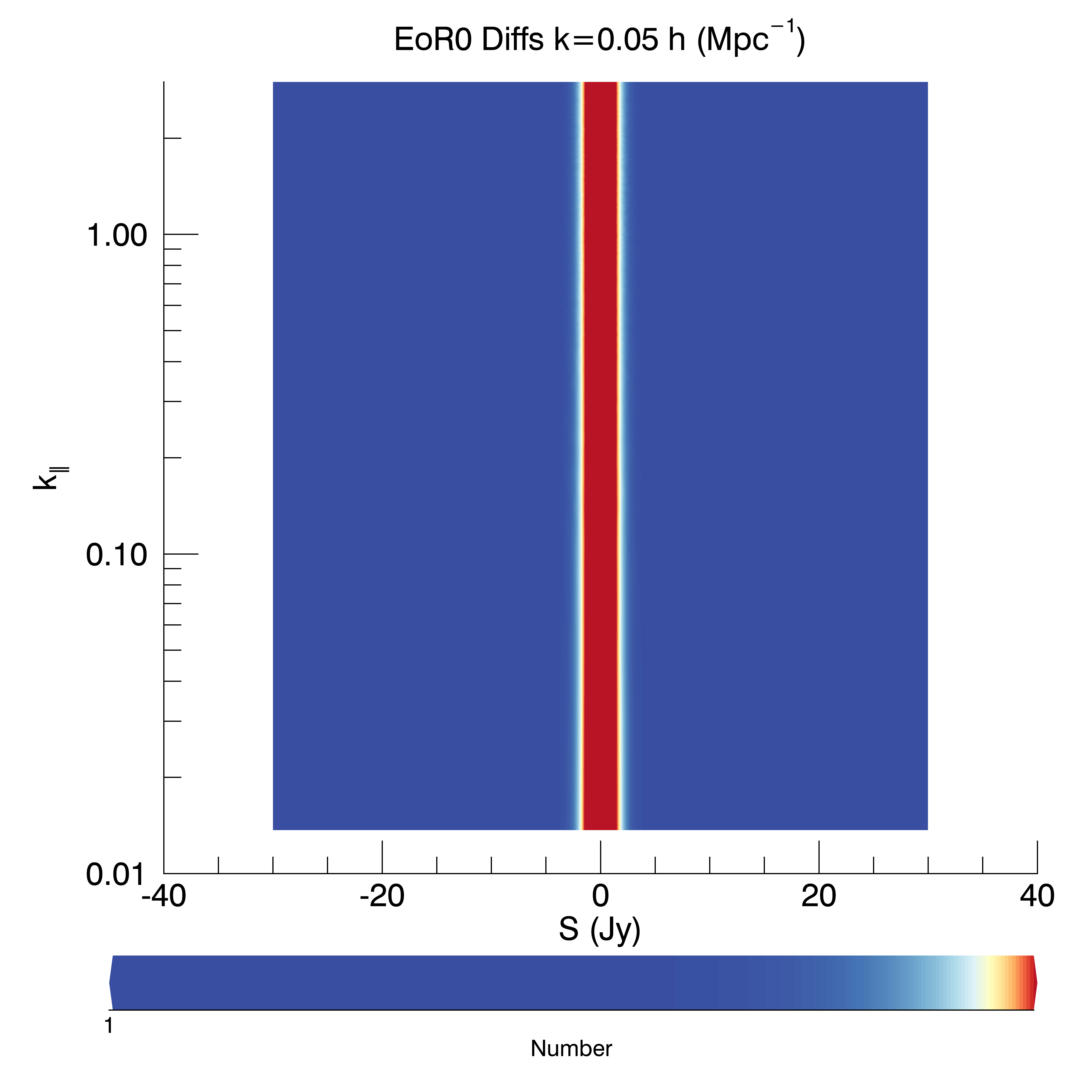}{(d)}
\caption{Heat maps {of the real parts} of visibilities (abscissa: flux density; ordinate: $k_\parallel$) of density estimates from 13 hours of data. (a, b) Difference visibilities (sum over adjacent interleaved time samples per baseline) at $k_\bot=0.005h$~Mpc$^{-1}$ for both fields. (c, d) Same, but for smaller scales, $k_\bot=0.05h$~Mpc$^{-1}$. All colour bars have the same scale, {corresponding to 0.1\% (white), 1\% (yellow), 10\% (orange), 50\% (red) of maximum.}}
\label{fig:teststat}
\end{figure*}

\begin{figure*}
\includegraphics[width=0.45\textwidth]{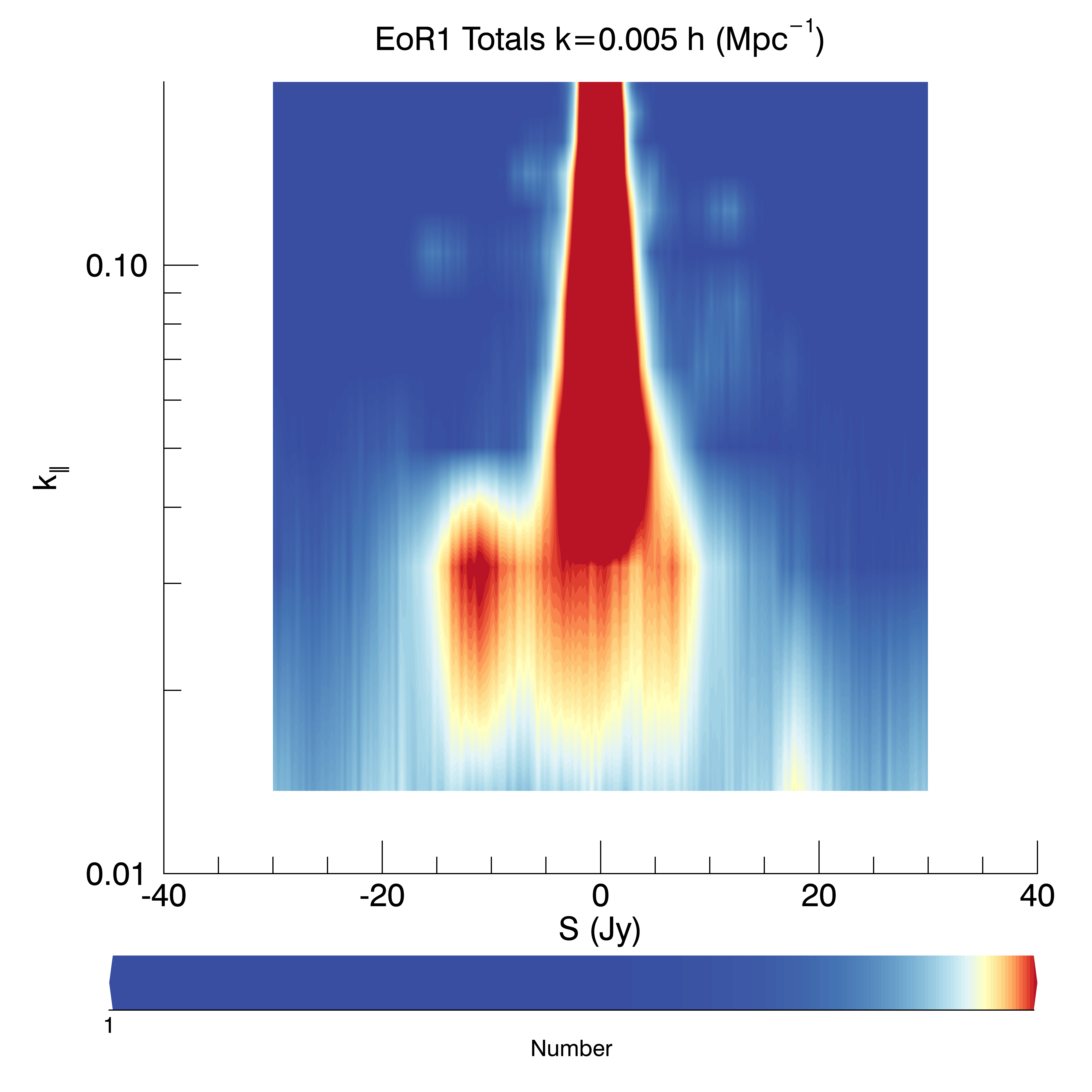}
\includegraphics[width=0.45\textwidth]{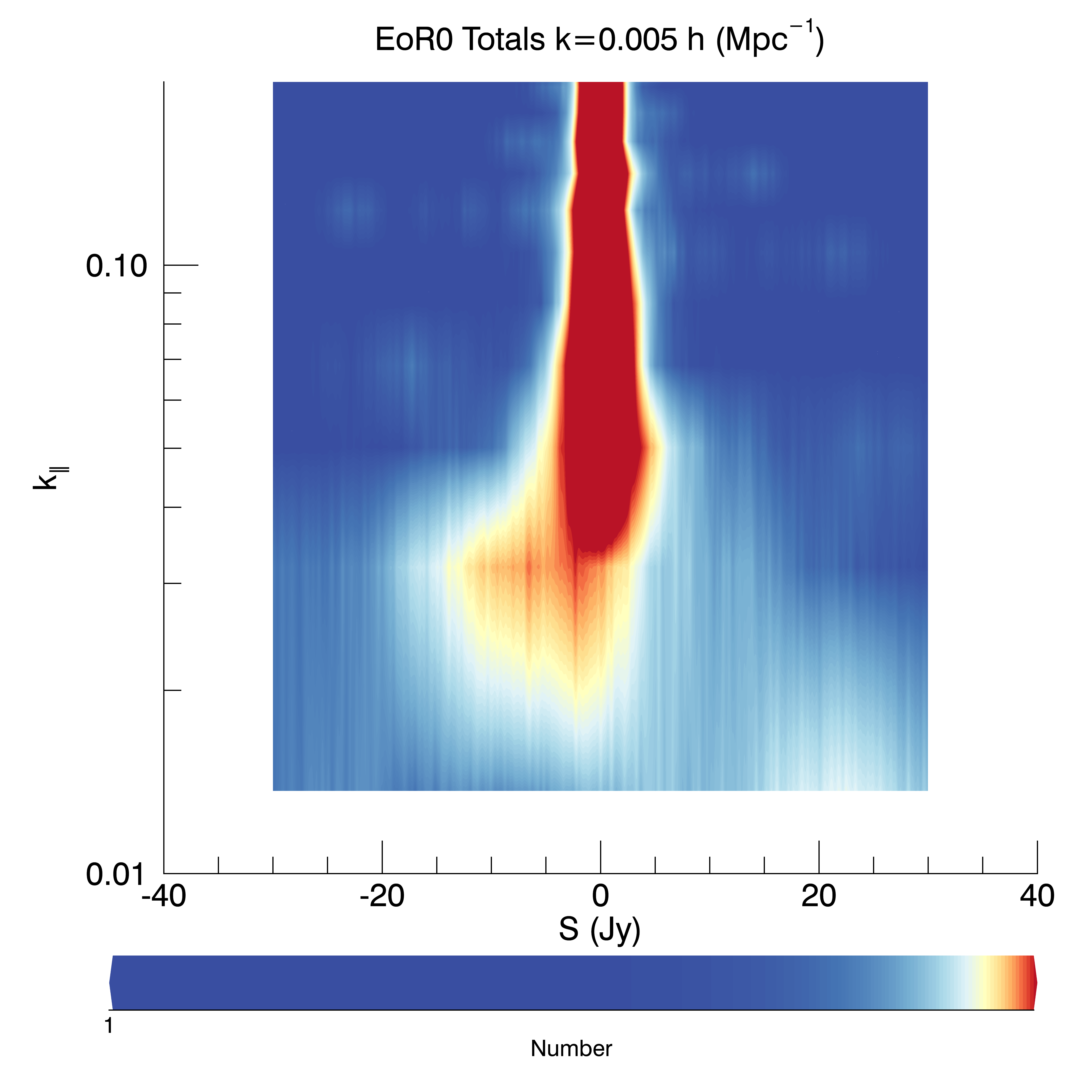}\\
\caption{Heat maps {of the real parts} of totals visibilities for both fields and the lower portion of $k_\parallel$ showing the structure on large scales. All colour bars have the same scale, {corresponding to 0.1\% (white), 1\% (yellow), 10\% (orange), 50\% (red) of maximum.}}
\label{fig:zoom}
\end{figure*}

Taking cuts through the parameter space, the histograms can be stacked together to be represented as contour plots. Figures \ref{fig:teststat2} and \ref{fig:teststat} display heat maps for two angular scales, both observing fields, and totals and difference visibilities (real part). The colour bars are the same for each set of panels, and represent the number of visibilities that contributed to the KDE distribution. The important features lie in the structure of the histograms, and not in the amplitudes, however the peak contour level is set to 1.2$\times$10$^{6}$ to capture the narrowest peak.
Figure \ref{fig:zoom} shows the lower $k_\parallel$ regions in more detail for the Totals.
At low $k_\bot-k_\parallel$ (large scales), the histograms are broad and structured, reducing to Gaussian-like for cleaner modes. The broad histograms at regular $k_\parallel$ intervals are due to regular missing spectral channels in the data, leading to harmonics of the $k_\parallel=0$ mode.

From these histograms, it is clear that the KDE yields additional information to what is available from gridding and squaring the visibilities, as is usual for a power spectrum analysis. The challenge is to understand how this information can be leveraged to better discriminate foregrounds from cosmological signal.

\begin{figure*}
\includegraphics[width=0.5\textwidth]{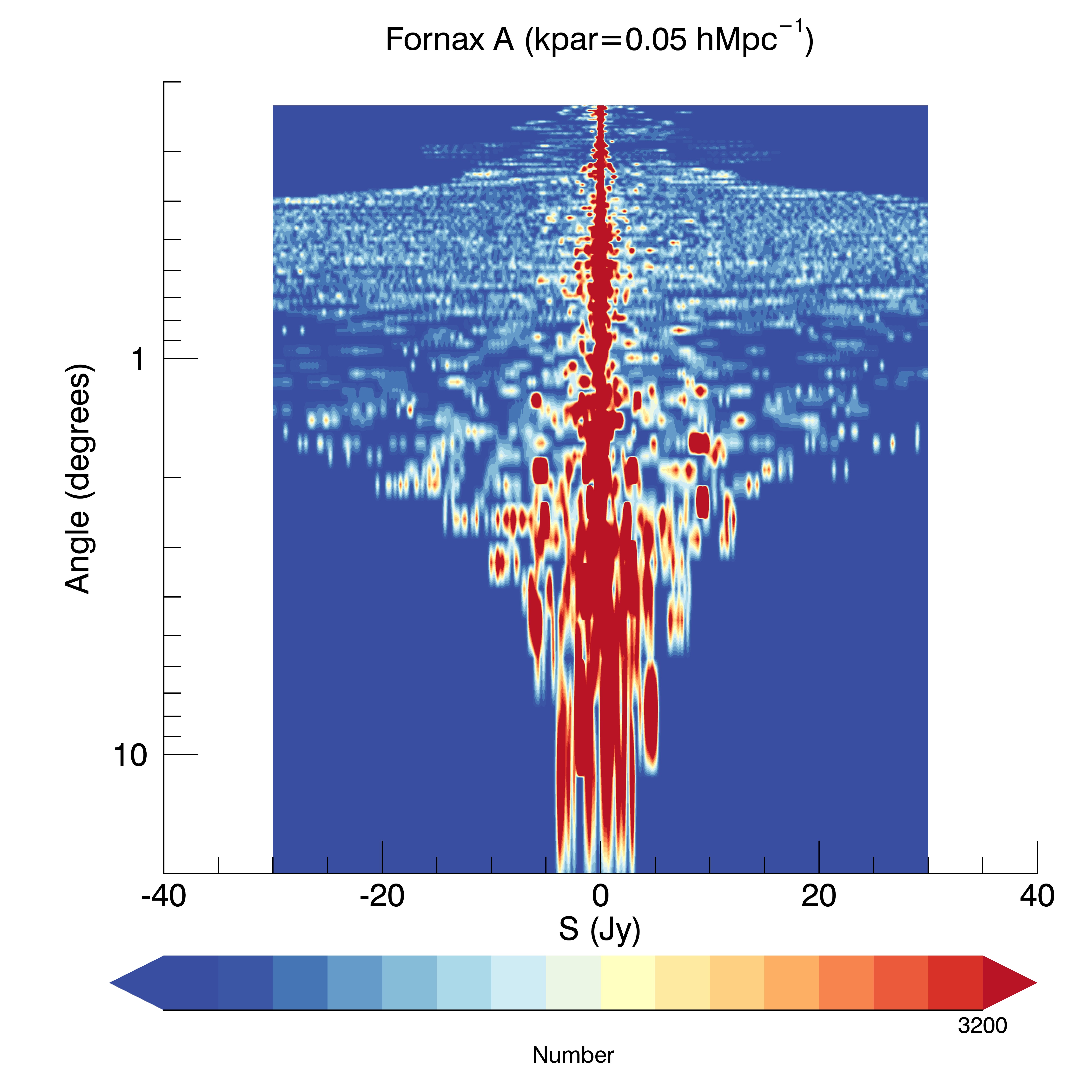}
\caption{Heat map for a shapelet-based model of Fornax A, as used in the MWA EoR calibration and peeling pipeline. Fornax A appears down the main lobe of the EoR1 field. The power is distributed across a range of scales, consistent with the structure of Fornax A, with the broadest distribution of measured flux densities at angular scales of 0.5--1.0 degree.}
\label{fig:forA}
\end{figure*}

{As a point of comparison, we compute the expected KDE histograms for a realistic model of Fornax A, a prominent extended, bright, structured source in the EoR1 observing field. Fornax A is the most prominent source in this field and its model is used in the EoR data calibration and source peeling steps. A shapelet-based model is constructed from MWA Phase I and Phase II extended array data \citep{line19}, with a simple constant spectral index across the band ($-0.8$). We expect the power to be concentrated at low $k_\parallel$. This model is currently used for our calibration. Details can be found in \citet{line19}. The model is fitted in image space, but because shapelet basis functions have known Fourier transforms, visibilities matching a zenith-pointed observation of the MWA can be directly calculated. No noise is added. These visibilities are then fed to the same KD-estimation software as for the real data. Figure \ref{fig:forA} shows a heat map of the distribution of power for $k_\parallel=0.05h$~Mpc$^{-1}$. The y-axis label has been converted from $k_\bot$ to angular degrees to show the concentration of power on 0.5--1.0 degree scales, consistent with the size of Fornax A.}
{Given the broad distribution of size scales, and the breadth of the distributions measured in Jansky, which exceed the datasets for the full unsubtracted Fornax A model, it is residuals from these sources that we expect to observe differently in each observing field.}

\subsection{Moments analysis}
\begin{figure*}
\includegraphics[width=0.85\textwidth]{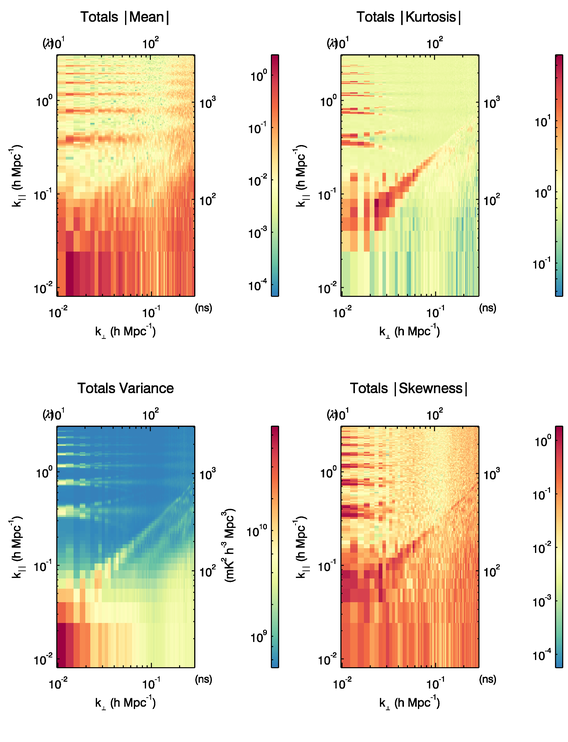}
\caption{EoR0 field moments using the KDE distributions from the real parts of the Totals visibilities. The third and fourth moment are dimensionless. {The horizon power shown diagonally at the interface of the EoR Window and foreground wedge shows prominently in the skewness and kurtosis, where power from large angle to the phase centre imprints highly non-Gaussian structure.}}
\label{fig:eor0_moments}
\end{figure*}

\begin{figure*}
\includegraphics[width=0.85\textwidth]{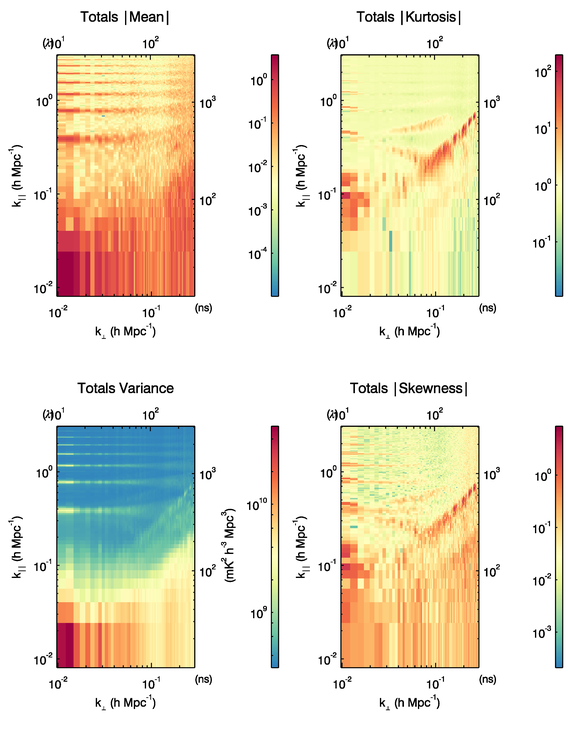}
\caption{EoR1 field moments using the KDE distributions from the Totals visibilities. The third and fourth moment are dimensionless. {The horizon signal in the skewness and kurtosis is less prominent than in the EoR0, where the Galactic Centre resides at the horizon.}}
\label{fig:eor1_moments}
\end{figure*}

Figures \ref{fig:eor0_moments} and \ref{fig:eor1_moments} display the first four moments for the totals visibilities in the EoR0 and EoR1 fields, respectively. The mean (first), skewness (third) and kurtosis (fourth) moments are displayed as their absolute values.
There are notable features to these results:
\begin{itemize}
    \item Non-zero mean values are associated with foreground-dominated modes
    \item Skewness exceeding the maximum value expected for cosmological signal is associated with foreground-dominated modes, but is most prominent at the edge of the foreground wedge
    \item Excess kurtosis appears at the edge of the foreground wedge, but is negligible in the main wedge
    \item The values and distributions differ between the two observing fields.
\end{itemize}
The significant skewness and kurtosis at the edge of the main wedge highlights the strong non-Gaussianity of the foregrounds at the observing horizon where foregrounds are adding a strong tail to both sides of the KDE histograms.
Most interestingly, the existence of non-zero means, excess skewness and non-zero excess kurtosis in modes known to be associated with foreground contamination suggests that this information may be used to weight modes in the final 1D spherically-averaged power spectrum. The differences between the two observing fields can further be used to tailor this.

\begin{figure*}
\includegraphics[width=0.85\textwidth]{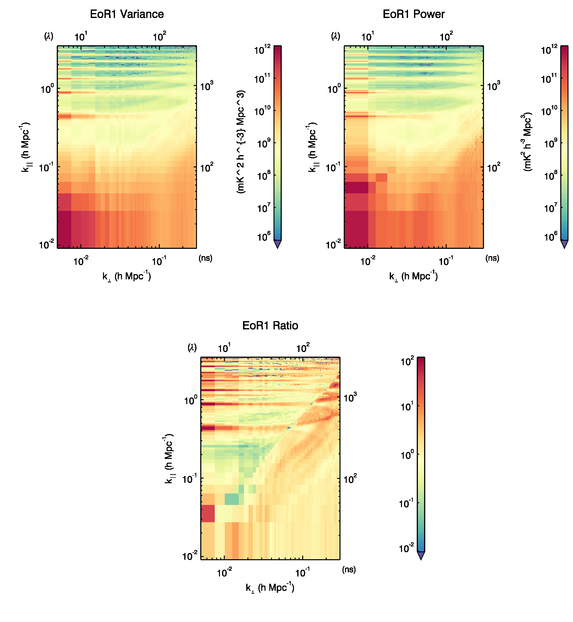}
\caption{Variance (second moment) and power spectrum for the EoR1 field (top), and ratio between the two (bottom). Computing the second moment from the histograms improves the performance within the EoR window.}
\label{fig:eor1_ratio}
\end{figure*}

The variance merits particular scrutiny because it is equivalent to the power spectrum when the data are Gaussian. Unlike the delay space estimator, where all of the structure of the visibility-derived KDE histograms gets included in the power spectrum estimate, the variance of the histograms only contains the second moment, removing any non-zero mean and ignoring higher-order moments. Therefore, the variance can be used as a clean, Gaussian version of the power spectrum. The variance for the EoR0 field exhibits strong harmonic lines parallel to the wedge, indicative of the power from the Galaxy near the horizon during these observations combined with the primary beam response function. Figure \ref{fig:eor1_ratio} contains the variance for EoR1 (scaled to cosmological units), the delay spectrum power spectrum estimate (reproduced from Figure \ref{fig:eor1_ps}), and the ratio of the two.
The ratio shows that the variance is able to provide a cleaner estimate in the EoR Window, where low-level non-Gaussian foregrounds are excised. This is a region of parameter space of high current interest in observational EoR experiments, where reducing leaked foregrounds can have substantial gains.

\begin{figure*}
\includegraphics[width=0.85\textwidth]{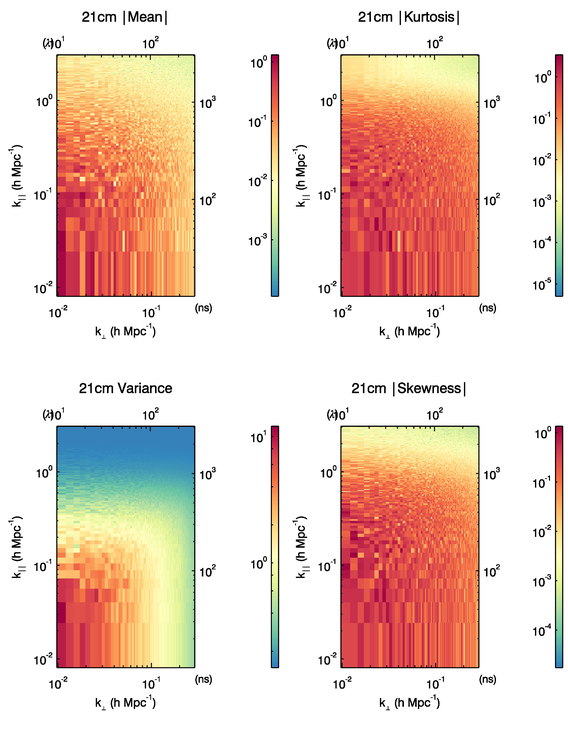}
\caption{21~cm moments extracted from a 21cmFAST simulation of a brightness temperature cube centred at $z$=8.28. The cube is Fourier Transformed and gridded to produce model visibilities, which are then fed into the same KDE software as the data. Aside from the variance, which mimics the power spectrum, the other moments show little structure.}
\label{fig:21cm_moments}
\end{figure*}

\begin{figure*}
\includegraphics[width=0.85\textwidth]{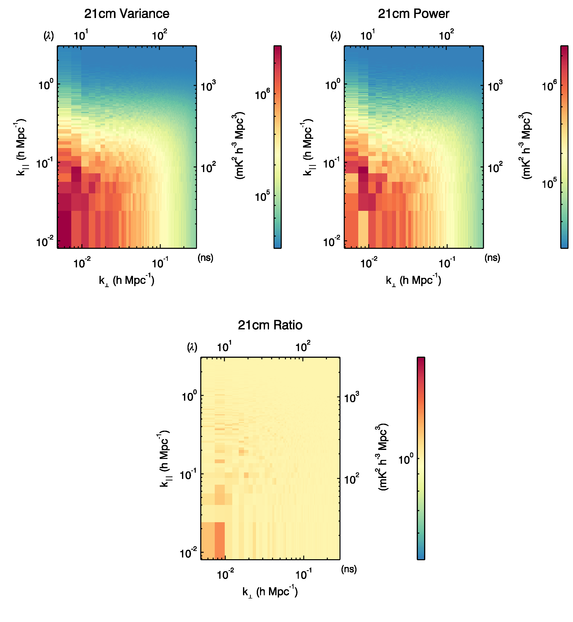}
\caption{Variance (second moment) and power spectrum for the 21~cm simulation (top), and ratio between the two (bottom), for comparison with Figure \ref{fig:eor1_ratio}. For this simple simulation, where the data are highly Gaussian, there is little difference.}
\label{fig:21cm_ratio}
\end{figure*}

For comparison, the first four moments are computed for mock visibilities derived from a 21cmFAST simulation \citep{mesinger11} at $z=8.28$ (\textsc{`delta\_T\_v3\_no\_halos\_zstart005.00000\_zend009.56801 \_FLIPBOXES0\_1024\_1600Mpc\_lighttravel'} available from http://homepage.sns.it/mesinger/EOS.html, Figure \ref{fig:21cm_moments}). {The simulation yields a cube of angular size 14~degrees (half the MWA beam), which is tapered with a Blackman-Harris window function, sampled at the same spectral resolution as the data and Fourier Transformed in angular scales to extract a $uv\nu$ cube. These model visibilities are converted to Jansky units and then gridded to the same resolution as the data, and fed through the same KDE pipeline. They therefore represent a `true' 21~cm signal, whereby the effects of the instrument are not included, and are therefore indicative plots of the moments for the 21~cm signal.} The mean, skewness and kurtosis all have values of $\simeq$~0.01--0.4 across the parameter space. The 21cmFAST boxes are small compared with the MWA beam size, yielding fewer samples to accurately estimate the distribution function. This leads to over-estimates of the moments due to the coarseness of the histograms, and so these plots should be taken as showing upper limits to the cosmological moments. 
{As a comparison to Figure \ref{fig:eor1_ratio}, where the ratio of the power spectrum to the second moment for the data was demonstrated to show structured differences, we compute the same ratio for the simulated 21~cm data. Figure \ref{fig:21cm_ratio} shows the same ratios.}
{There is little difference observed between the two approaches. This is partly owing to the high degree of Gaussianity in the data extracted from 21cmFAST simulations, but also reflects the smoothness of the expected signal when devoid of instrumental effects. These results are encouraging, but not definitive, to demonstrate that use of the second moment can be advantageous without signal loss.}

\subsection{Phase of moments}
\begin{figure*}
\includegraphics[width=0.75\textwidth]{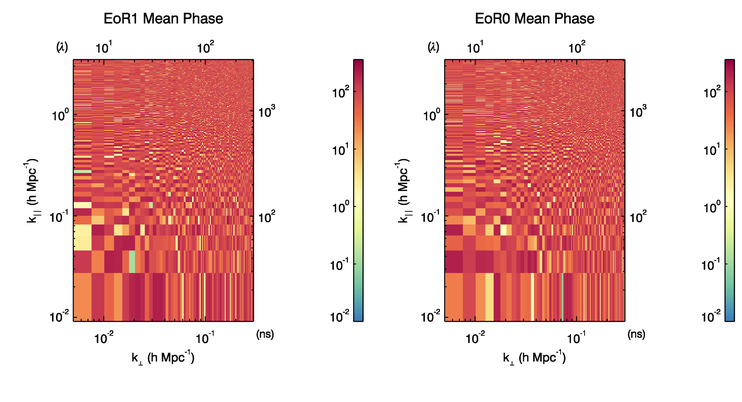}{(a)}\\
\includegraphics[width=0.75\textwidth]{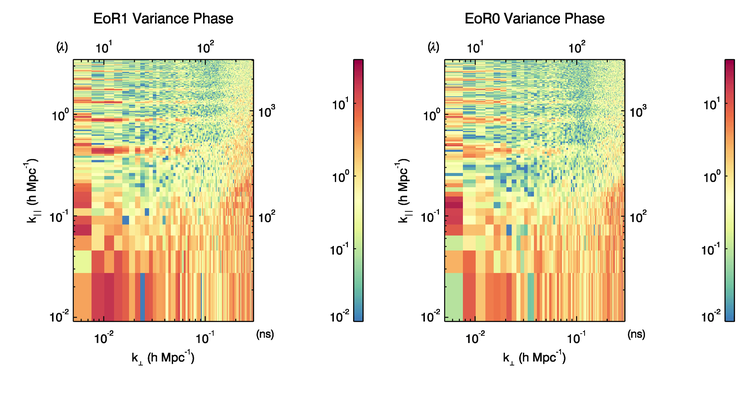}{(b)}
\caption{Phase for the first two moments and each observing field (measured in degrees). A phase close to zero indicates equal values for the real and imaginary components. The mean shows consistent offsets for the real and imaginary components, due to that statistic using all data equally, whereas the variance shows departures, consistent with non-stochastic foreground sources away from phase centre having a defined and different real and imaginary part.}
\label{fig:phase_1}
\end{figure*}

\begin{figure*}
\includegraphics[width=0.75\textwidth]{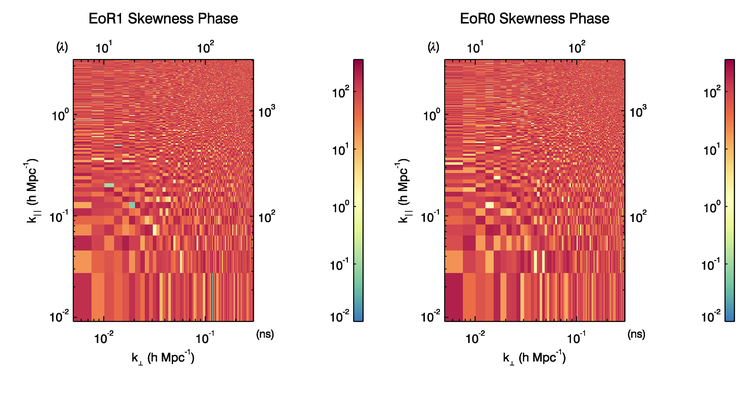}{(a)}\\
\includegraphics[width=0.75\textwidth]{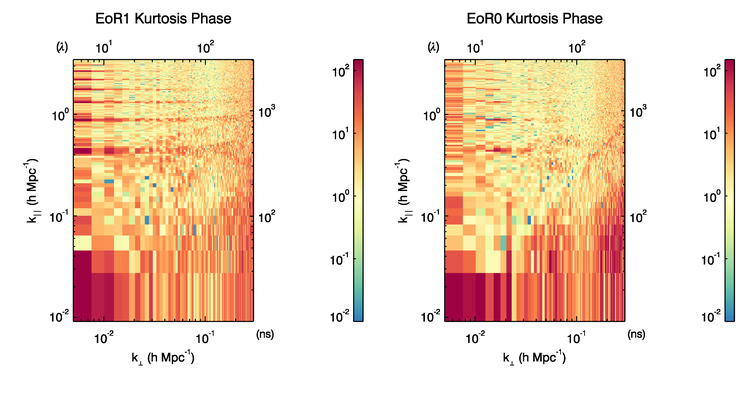}{(b)}
\caption{Phase for the third and fourth moments and each observing field (measured in degrees). A phase close to zero indicates equal values for the real and imaginary components.}
\label{fig:phase_2}
\end{figure*}

While the gridding of visibilities through the KDE cannot retain phase information, the effective phase of the moments can be extracted for each $k_\bot-k_\parallel$ mode, where the phase here encodes the relative angle between the real and imaginary component moment values. Figures \ref{fig:phase_1} and \ref{fig:phase_2} display the phase for each of the four moments and each observing field (measured in degrees). A phase close to zero indicates equal values for the real and imaginary components.

Strong non-zero phase values indicate differences between the two components and are suggestive of foreground contamination (for modes with sufficient samples). Values close to 90~degrees across the full parameter space for the mean values and skewness reflect that most modes have equal values in both the real and imaginary components, but opposite sign (i.e., they exhibit near-uniform amplitudes of $\sim$90~degrees). Foreground contamination is most obvious in the variance and kurtosis. It is difficult to see whether the phase adds more information than is available in the amplitudes of the moments alone. 

\subsection{Earth Mover's Distance}
\begin{figure*}
\includegraphics[width=0.75\textwidth]{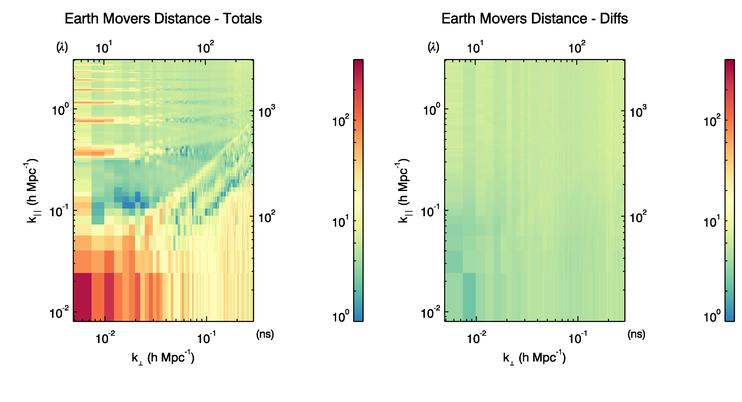}
\caption{Earth Mover's distance between histogrammed totals (left) and differences (right) visibilities from the two observing fields. Clearly the foreground-dominated regions are prominent with this metric. {The apparent lower value of the totals EMD compared with the differences EMD at $k_\bot \simeq 0.02h$~Mpc$^{-1}$, $k_\parallel \simeq 0.11h$~Mpc$^{-1}$ is consistent with the expected variation in the expected value, which is of comparable amplitude.}}
\label{fig:emd}
\end{figure*}

To fully utilize the differences in the statistical properties of foregrounds in different patches of the sky, the Earth Mover's Distance quantifies the degree of dis-similarity between the KDE distributions. Figure \ref{fig:emd} displays the EMD between EoR0 and EoR1 for the totals and differences visibilities.
The differences EMDs are smooth across the parameter space, with values of 2--5, which can be compared with the expectations from Equation \ref{eqn:emd}, where $\langle{\rm EMD}\rangle \simeq 1.5$. The totals EMDs exhibits clear structure where the foregrounds differ between the two fields. This is particularly prominent at the edge of the wedge, where large-scale foregrounds close to the horizon differ (EoR0 contains horizon emission from the Galaxy, whereas EoR1 contains the extended radio source Fornax A near the edge of the primary beam). The EMD encodes the regions of the power spectrum where foregrounds are prominent and can be excised. It is not strong in the EoR Window, where one might hope that it could be used for foreground discrimination and weighting. The apparent lower value of the totals EMD compared with the differences EMD at $k_\bot \simeq 0.02h$~Mpc$^{-1}$, $k_\parallel \simeq 0.11h$~Mpc$^{-1}$ is consistent with the expected variation (uncertainty) in the expected value, which is of comparable amplitude, i.e., $\Delta$EMD$\simeq 1.5$.

\subsection{Spherically-averaged power spectrum}
The previous sections have explored different statistics available from the KDEs, with a view to using the additional information to discriminate foregrounds from EoR signal. In general, the EoR Window is the region of parameter space of most interest to clean, because this region shows only moderate foreground contamination, but also contains a relatively large expected cosmological signal.

\begin{figure*}
\includegraphics[width=0.75\textwidth]{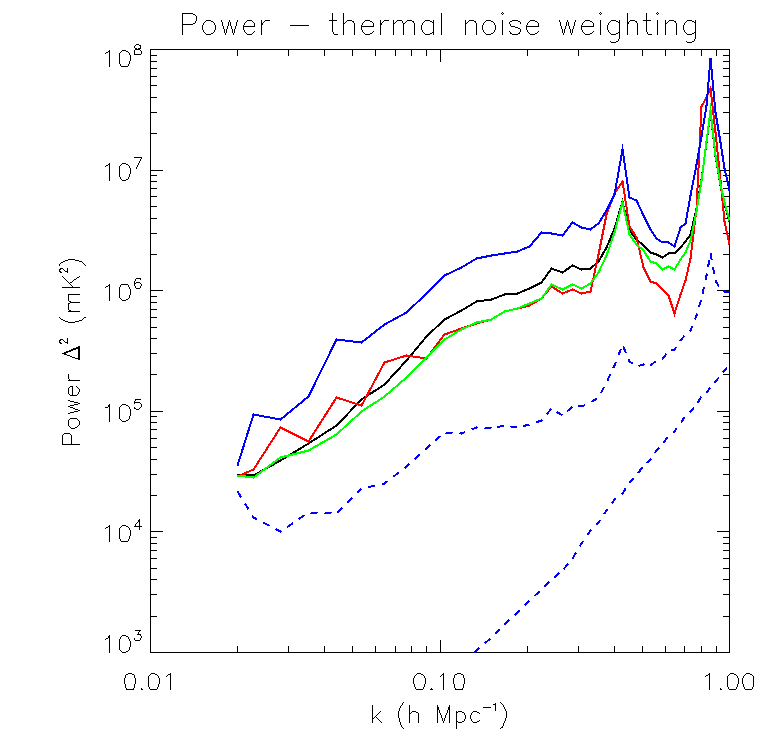}
\caption{Spherically-averaged dimensionless power spectra using thermal noise-only weighting, but leveraging differences between the two observing fields and the KDE moments analysis. (Blue) Regular delay spectrum; (black) EoR0 field second moment; (red) EoR1 field second moment; (green) cleaned second moment using differences between EoR0 and EoR1; (dashed blue) thermal noise and thermal noise$+$sample variance. {Most prominently, use of the second moment alone yields improvement, with varying performance of the different fields at different $k$.}}
\label{fig:1d_out}
\end{figure*}

The results of Figure \ref{fig:eor1_ratio} suggest that use of the second moment alone (compared with the typical CHIPS-like estimator) can provide improvement in the EoR Window. Motivated by this, we first compute the dimensionless spherically-averaged power spectrum using only thermal noise weighting, but for different fields. Figure \ref{fig:1d_out} displays the power spectra using thermal noise-only weighting, but leveraging differences between the two observing fields and the KDE moments analysis. To do this, we {take the KDE histograms for a given $k_\bot-k_\parallel$ cell for each of the observing fields and form the narrowest histogram from the combined datasets, thereby excising measurements that are not consistent between the two fields. While this appears to be a severe cut, in practise, this only removes portions of each histogram that are at the extrema. After excision, both fields are left with knowledge of the other field's information, such that differences in the foreground statistics have been identified and reduced.} The KDE histograms such that the second moment is computed from:
\begin{equation}
    f_{\rm clean} = {\rm min}\left[ f_{\rm EoR0} , f_{\rm EoR1} \right].
\end{equation}
The figure displays (blue) Regular delay spectrum; (black) EoR0 field second moment; (red) EoR1 field second moment; (green) cleaned second moment using differences between EoR0 and EoR1; (dashed blue) thermal noise and thermal noise$+$sample variance.
{The excision of inconsistent parts of the histograms (green) yields some small improvement at low $k$, but is less useful than the second moment alone at higher $k$ values.} These results demonstrate that accessing the second moment alone can remove field-unique foregrounds (and any non-Gaussian cosmological signal), and that use of information from two observing fields offers moderate improvement. The former yields an improvement by a factor of 2--3 in power for $k < 0.3h$~Mpc$^{-1}$, while the latter yields factors of 1--2. The sample variance curve also demonstrates that these modes have well-sampled statistics in order for the KDE to yield reliable distributions.

\begin{figure*}
\includegraphics[width=0.75\textwidth]{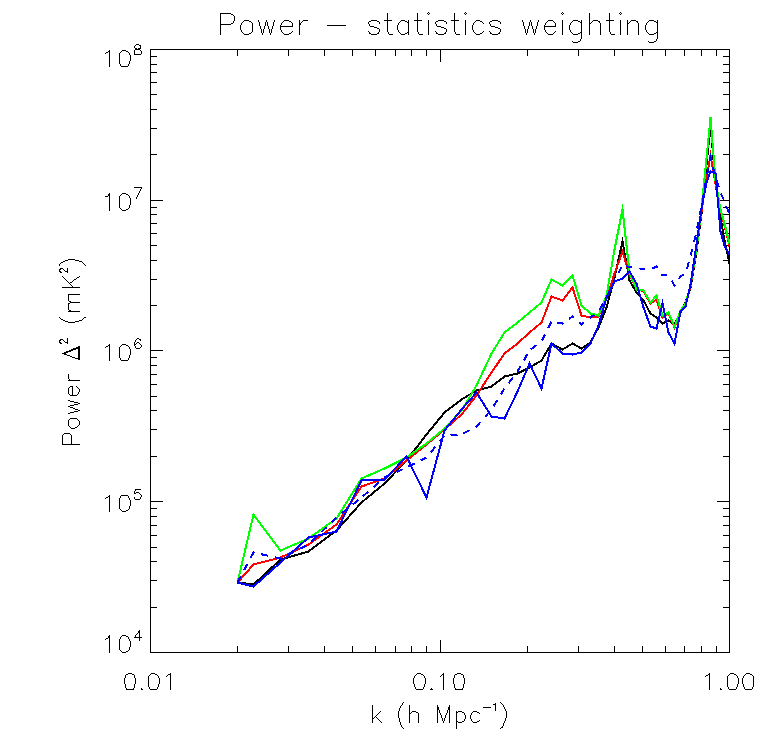}
\caption{Spherically-averaged dimensionless power spectra using weightings based on statistics, applied to the cleaned second moment using differences between EoR0 and EoR1 (green line from Figure \ref{fig:1d_out}). (Black) thermal noise weighting; (red) weighting including thermal, non-zero mean, skewness and kurtosis; (green) weighting including thermal, non-zero kurtosis; (blue) weighting including phase of second moment; (blue dashed) weighting including thermal and EMD.}
\label{fig:1d_out_stats}
\end{figure*}

We now focus on using the cleanest second moments (the histograms cleaned of {large flux density values} that are inconsistent between the two observing fields and therefore hypothesised to be due to foregrounds), \textit{and} change the weighting to include some of the other statistics explored in this work: first, third and fourth moments, EMD and phase of the second moment. For example, weighting by thermal noise and the EMD yields weights for cell $i$:
\begin{equation}
    W_i = \frac{1}{\frac{1}{W_{\rm therm}} + \alpha|{\rm EMD}|},
\end{equation}
where $\alpha$ scales the absolute EMD value to be of comparable amplitude to the thermal noise weights.
Figure \ref{fig:1d_out_stats} shows five curves with different weightings: (black) thermal noise weighting; (red) weighting including thermal, non-zero mean, skewness and kurtosis; (green) weighting including thermal, non-zero kurtosis; (blue) weighting including phase of second moment; (blue dashed) weighting including thermal and EMD.
There are moderate differences in the range $k = 0.1-0.4h$~Mpc$^{-1}$, but the main region of interest ($k \sim 0.1h$~Mpc$^{-1}$) is not impacted by the different weighting. These statistics are therefore not providing significant \textit{additional} information, {suggesting that use of the second moment alone captures most of the improvement. This can be understood as follows: the second moment, by construction, ignores any low-level tails, non-zero mean or large-Janksy values in the histogram. These are the measurements that are mathematically favoured in calculation of the kurtosis and skewness, and therefore use of these statistics in the weighting is only down-weighting cells that are already unfavoured. Similarly, the EMD shows large values in the same parts of parameter space, because it is capturing information about measurements against which the second moment is robust}. Similar results are found when the thermal noise is removed from the weighting entirely, and the additional statistics only are used. Use of the phase of the second moment offers some advantages at select $k$ modes. {This can be seen in the phase of the second moment (Figure \ref{fig:phase_1}, lower) in the $k_\bot$=0.01-0.02$h$Mpc$^{-1}$, $k_\parallel$=0.1-0.2$h$Mpc$^{-1}$ region, where the phase is large further into the EoR Window than other statistics are showing contamination (cf Figure \ref{fig:emd}).}

We note that these data processed through the CHIPS estimator yield power estimates that are a factor of 2--4 better than the delay space estimator. This is a generic result because CHIPS correctly handles the beam gridding. A future KDE approach with a similar uv-gridded estimator could be expected to yield improvement again over the CHIPS estimator, because its improvement stems from its ability to clean non-Gaussian foregrounds that are otherwise captured in a normal visibility-squaring approach (a group to which CHIPS and the delay space estimator both belong).

\section{Discussion and conclusions}
The moments analysis, phase of moments and EMD are all shown to contain information about foregrounds and non-Gaussian components. The EMD also has the advantage of not assuming Gaussianity of the underlying signal, preserving the full cosmological distribution function of temperature fluctuations. However, in the spherically-averaged analysis, these statistics did not offer improvement when used in the weights applied to each cell before spherical averaging.

Instead, the ability to directly access the second moment through the KDE analysis allowed for factors of 2$-$3 improvement in power in the region of interesting modes. This is due to the second moment being able to ignore large-flux density non-Gaussian foreground-contaminated data, unlike the typical power spectrum estimator, which incorporates all data blindly. However, this also destroys non-Gaussian 21~cm signal power.

In this work we emphasise the usefulness of comparing data from two independent observing fields as an avenue to robustly discriminate statistically-dissimilar foregrounds from statistically-similar cosmological signal, particularly when the avenues can be agnostic to the form of the EoR temperature fluctuation distribution function. The signal is expected to be non-Gaussian with a non-zero third moment (skewness), a factor ignored by power spectrum analysis (the variance is expected to contain most of the information, and so this choice is generally warranted for current experiments). Guided by the preliminary results shown here, in future work, additional statistics will be developed based on the form of the KDE distribution functions, which may yield more promising discriminators for foregrounds and cosmological signal.

In future, the KDE approach can be extended to apply to a uv-gridded power spectrum estimator, such as CHIPS, which is able to be used for cosmological purposes because it truly estimates power in $k_\bot-k_\parallel$. There is no methodological impediment to this approach, but it is computationally more expensive than the delay spectrum. In this work, the delay spectrum estimator yields results that are a factor of 2--4 worse than a CHIPS estimator, and we would therefore expect that a future gridded KDE approach may yield a factor of 2--3 improvement over the CHIPS estimate, thereby making it relevant for current experiments.

\section*{Acknowledgements}
We thank the anonoymous referee for the many comments that have clarified the methodology and improved the exposition.
This research was supported by the Australian Research Council Centre of Excellence for All Sky Astrophysics in 3 Dimensions (ASTRO 3D), through project number CE170100013. CMT is supported by an ARC Future Fellowship under grant FT180100196.
The International Centre for Radio Astronomy Research (ICRAR) is a Joint Venture of Curtin University and The University of Western Australia, funded by the Western Australian State government.
The MWA Phase II upgrade project was supported by Australian Research Council LIEF grant LE160100031 and the Dunlap Institute for Astronomy and Astrophysics at the University of Toronto.
This scientific work makes use of the Murchison Radio-astronomy Observatory, operated by CSIRO. We acknowledge the Wajarri Yamatji people as the traditional owners of the Observatory site. Support for the operation of the MWA is provided by the Australian Government (NCRIS), under a contract to Curtin University administered by Astronomy Australia Limited. We acknowledge the Pawsey Supercomputing Centre which is supported by the Western Australian and Australian Governments.

\bibliographystyle{mnras}
\bibliography{mnras_template} 

\bsp	
\label{lastpage}
\end{document}